\documentclass[iop, twocolappendix]{emulateapj}

\usepackage{amsmath, amssymb}
\usepackage{graphicx}
\usepackage{multirow}
\usepackage{natbib}
\usepackage{url}
\usepackage{aas_macros}
\usepackage{my_macros}

\usepackage[backref, breaklinks, colorlinks, citecolor=blue, linkcolor=blue, urlcolor=blue]{hyperref}
\usepackage[all]{hypcap}

\newcommand{\figurewidth}{0.99}
\newcommand{\figureheight}{0.2}

\begin{document}

\journalinfo{The Astrophysical Journal, submitted}

\title{Satellite Dwarf Galaxies in a Hierarchical Universe: Infall Histories, Group Preprocessing, and Reionization}
\shorttitle{Satellite Infall Histories, Group Preprocessing, and Reionization}
\shortauthors{Wetzel, Deason \& Garrison-Kimmel}

\author{Andrew R. Wetzel \altaffilmark{1, 2}}
\altaffiltext{1}{TAPIR, California Institute of Technology, Pasadena, CA, USA}
\altaffiltext{2}{Carnegie Observatories, Pasadena, CA, USA}
 
\author{Alis J. Deason \altaffilmark{3, 4}}
\altaffiltext{3}{Department of Astronomy and Astrophysics, University of California, Santa Cruz, CA, USA}
\altaffiltext{4}{Hubble Fellow}

\author{Shea Garrison-Kimmel \altaffilmark{5}}
\altaffiltext{5}{Center for Cosmology, Department of Physics and Astronomy, University of California, Irvine, CA, USA}

\begin{abstract}
In the Local Group (LG), almost all satellite dwarf galaxies that are within the virial radius of the Milky Way (MW) and Andromeda (M31) exhibit strong environmental influence.
The orbital histories of these satellites provide the key to understanding the role of the MW/M31 halo, lower-mass groups, and cosmic reionization on the evolution of dwarf galaxies.
We examine the virial-infall histories of satellites with $\mstar=10^{3-9}\msun$ using the ELVIS suite of cosmological zoom-in dissipationless simulations of 48 MW/M31-like halos.
Satellites at $z=0$ fell into the MW/M31 halos typically $5-8\gyr$ ago at $z=0.5-1$.
However, they first fell into any host halo typically $7-10\gyr$ ago at $z=0.7-1.5$.
This difference arises because many satellites experienced ``group preprocessing'' in another host halo, typically of $\mvir \sim 10^{10-12}\msun$, before falling into the MW/M31 halos.
Satellites with lower mass and/or those closer to the MW/M31 fell in earlier and are more likely to have experienced group preprocessing; half of all satellites with $\mstar<10^6\msun$ were preprocessed in a group.
Infalling groups also drive most satellite-satellite mergers within the MW/M31 halos.
Finally, \textit{none} of the surviving satellites at $z=0$ were within the virial radius of their MW/M31 halo during reionization ($z>6$), and only $<4\%$ were satellites of any other host halo during reionization.
Thus, effects of cosmic reionization versus host-halo environment on the formation histories of surviving dwarf galaxies in the LG occurred at distinct epochs, separated typically by $2-4\gyr$, so they are separable theoretically and, in principle, observationally.
\end{abstract}


\section{Introduction}

Galaxies in dense environments are more likely to have suppressed (quiescent) star-formation rates, more elliptical/spheroidal morphologies, and less cold gas in/around them than galaxies of similar stellar mass, $\mstar$, in less dense environments.
While such environmental effects long have been studied for galaxies in massive galaxy groups and clusters \citep[for example,][for review]{Oemler1974, Dressler1980, DresslerGunn1983, Balogh1997, BlantonMoustakas2009}, the observed effects on the dwarf galaxies in the Local Group (LG), in particular, the satellites within the host halos of the Milky Way (MW) and M31, are even stronger \citep[for example,][]{Mateo1998, McConnachie2012, Phillips2014, SlaterBell2014, Spekkens2014}.

Specifically, the galaxies around the Milky Way (MW) and Andromeda (M31) show a strikingly sharp transition in their properties within $\approx 300 \kpc$, corresponding to the virial radii, $\rvir$, of the halos of the MW and M31 for $\mvir \approx 10 ^ {12} \msun$ \citep[for example,][]{Deason2012, VanDerMarel2012, BoylanKolchin2013}.
Within this distance, galaxies transition from (1) having irregular to elliptical/spheroidal morphologies, (2) having most of their baryonic mass in cold atomic/molecular gas to having little-to-no detectible cold gas, and (3) being actively star-forming to quiescent \citep[][and references therein]{McConnachie2012}.
This environmental transition of the population is almost complete, with just a few exceptions.
Four gas-rich, star-forming, irregular galaxies persist within the halos of the MW (the LMC and SMC) and M31 (LGS 3 and IC 10).
However, the LMC and SMC are likely on their first infall \citep{Besla2007, Kallivayalil2013}, and given their distances to M31, LGS 3 and IC 10 may be as well.
Furthermore, 4 - 5 gas-poor, quiescent, spheroidal galaxies exist beyond the halos of either the MW and M31: Cetus, Tucana, KKR 25, KKs 3 \citep{Karachentsev2015} and possibly Andromeda XVIII.
While the radial velocities of Cetus and Tucana imply that they likely orbited within the MW halo \citep{Teyssier2012}, KKR 25 and KKs 3 are much more distant at $\approx 2 \mpc$.
The fact that almost all of the satellite galaxies within the MW/M31 halos show such strong environmental effects is particularly striking given that, other than KKR 25 and KKs 3, all known galaxies at $\mstar < 10 ^ 9 \msun$ that are isolated (not within $1500 \kpc$ of a more massive galaxy, and thus not strongly influenced by environmental effects) are actively star-forming \citep{Geha2012}.
Thus, the MW and M31 halos exert the strongest environmental influence on their galaxy populations of any observed systems, making the LG one of the most compelling laboratories to study environmental effects on galaxy evolution.

Several environmental processes within a host halo regulate the gas content, star formation, morphology, and eventual tidal disruption of satellite galaxies.
Gravitationally, the strong tidal forces of the host halo will strip mass from the satellite (subhalo) from the outside-in \citep{Dekel2003, Diemand2007, WetzelWhite2010}.
In addition, the dense collection of satellites within a host halo can drive impulsive gravitational interactions with each other \citep{FaroukiShapiro1981, Moore1998}, and satellites can merge with one another \citep{Angulo2009, Wetzel2009a, Wetzel2009b, Deason2014a}.
Moreover, tidal shocking and resonant interactions with the host's galactic disk can lead to particularly efficient morphological evolution, coring, stripping, and disruption \citep{Mayer2001, DOnghia2010, Zolotov2012}.
Hydrodynamically, if the host halo contains thermalized hot gas, this can strip and heat the extended gas from the orbiting satellite subhalo \citep{Balogh2000, McCarthy2008}, leading to reduced gas cooling/accretion into the satellite's disk \citep{Larson1980}.
More drastically, given a sufficiently high density of hot gas and high orbital velocity, ram-pressure can strip cold gas directly from the satellite's disk \citep{GunnGott1972, Abadi1999, Mayer2006, Chung2009, Tonnesen2009}.
Furthermore, feedback from stars and/or AGN within the satellites can drive galactic winds that can enable these environmental process to operate even more efficiently \citep[for example,][]{BaheMcCarthy2015}.

Understanding the relative efficiency of the above environmental processes, including the timescales over which they have operated, requires understanding in detail the orbital and virial-infall histories of the current satellite population in the context of the hierarchical structure formation of $\Lambda$CDM.
While some authors examined the virial-infall times of LG-like satellites in cosmological settings \citep[for example,][]{Lux2010, Rocha2012}, such works used cosmological zoom-in simulations of one or two MW-like halos, which does not model the environment of the MW/M31 pair in the LG or allow for good statistics.
In addition, hierarchical growth means that many satellites may have been in a group before they fell into the MW/M31 halos.
Environmental processing in such groups could help to explain the high efficiency and near completeness of environmental effects on the satellite population.
Several authors explored the importance of this ``group preprocessing'' on satellites within massive groups and clusters \citep[for example,][]{ZabludoffMulchaey1998, McGee2009, Hou2014}.
In particular, \citet{Wetzel2013} found that group preprocessing alone largely can account for the fact that satellites in more massive groups/clusters are more likely to be quiescent.
However, on mass scales of MW/M31 halos, the impact of group preprocessing on dwarf galaxies remains largely unexplored.
Using a cosmological zoom-in simulation of a single MW-like halo \citet{LiHelmi2008} found that $\sim 1 / 3$ of satellites fell in as part of a group, and using the two Via Lactea simulations, \citet{SlaterBell2013} similarly found that many satellites are organized into small groups with correlated infall.

If some of the satellites in the MW/M31 halos fell in as part of a group, this would have several implications for their subsequent evolution and spatial distribution.
For instance, group infall could have caused many of the strong associated in phase-space coordinates between the observed satellites (and streams) in the LG \citep{LyndenBellLyndenBell1995, DOnghiaLake2008, Klimentowski2010}, including the disk-like configurations of satellites around the MW and M31 \citep[for example,][]{Libeskind2005, Lovell2011, Fattahi2013, Ibata2013, PawlowskiKroupa2014}.
Furthermore, group infall could have driven mergers between satellites after infall \citep[for example,][]{Deason2014a}.

In addition to the above environmental processes that operate within a host halo, cosmic reionization may have had a lasting impact on formation histories of dwarf galaxies in the LG, by heating/removing the gas from low-mass halos whose virial temperatures were below that of the ultra-violet photoionization background, thus quenching star formation in the lowest-mass galaxies and leaving them as relics of reionization \citep{Bullock2000, Gnedin2000}.
Many ongoing observational efforts aim to use the current stellar populations of faint and ultra-faint satellites in the LG to test the impact that reionization may have had on their star-formation histories at $z \gtrsim 6$ \citep[for example,][]{Brown2014, Weisz2014b}.
However, a long-standing challenge for such studies is whether one can separate the effects of cosmic reionization from those of the host-halo environment on the formation histories of surviving satellite galaxies.

In this work, we examine the orbital and virial-infall histories of the current satellite galaxies of the LG in a fully cosmological and hierarchical context, including the impact of group preprocessing and implications for using such galaxies as probes of reionization.
Specifically, we will address the following questions for the satellites in the halos of the MW and M31:
\begin{enumerate}
\renewcommand{\labelenumi}{(\alph{enumi})}
\item When did they fall into the MW/M31 halo, and when did they first fall into any host halo?
\item What fraction were within their MW/M31 halo, or any other host halo, during cosmic reionization ($z > 6$)?
\item What fraction were in a group prior to falling into the MW/M31 halo?
\item What role does group infall play in driving mergers between satellites?
\end{enumerate}

\section{Numerical Methods}
\label{sec:method}

\subsection{Simulations}
\label{sec:simulation} 

To study the orbital histories of satellite dwarf galaxies, we use ELVIS (Exploring the Local Volume in Simulations), a suite of cosmological zoom-in $N$-body simulations that are targeted to modeling the LG \citep{GarrisonKimmel2014}.
ELVIS was run using \textsc{GADGET-3} and \textsc{GADGET-2} \citep{Springel2005e}, with initial conditions generated using \textsc{MUSIC} \citep{HahnAbel2011}, all with $\Lambda$CDM cosmology based on WMAP7 \citep{Larson2011}: $\sigma_8 = 0.801$, $\omegamatter = 0.266$, $\omegalambda = 0.734$, $n_s = 0.963$ and $h = 0.71$.

ELVIS contains 48 dark-matter halos of masses similar to the MW or M31 ($\mvir = 1.0 - 2.8 \times 10 ^ {12} \msun$) within a zoom-in volume of radius $\gtrsim 4\,\rvir$ of each halo at $z = 0$.
Half of these halos are located in zoom-in regions that were selected to contain a pair of halos that resemble the masses, distance, and relative velocity of the MW-M31 pair, while the other half are single isolated halos matched in masses to the paired ones.
These zoom-in regions are selected from a suite of simulations, each a cube with side length $70.4 \mpc$.
Within the zoom-in regions, the particle mass is $1.9 \times 10 ^ 5 M_\odot$ and the Plummer-equivalent force softening is $140 \pc$ (comoving at $z > 9$, physical at $z < 9$).
Additionally, three of the isolated halos were re-run at higher resolution, with particle mass of $2.4 \times 10 ^ 4 M_\odot$ and force softening of $70 \pc$.
Using these simulations, we checked that resolution does not significantly affect any results in this work.
See \citet{GarrisonKimmel2014} for more details on ELVIS.

Throughout this work, unless otherwise stated, we use the paired halos, which more accurately capture the environment, assembly history, and massive satellite population (LMC/M33-like satellites) of the LG.
More precisely, we use 10 halo pairs (20 halos total), ignoring the pairs Siegfried \& Roy and Serena \& Venus because they contain a massive halo within $1.2 \mpc$ that is not representative of the LG \citep{GarrisonKimmel2014}.
Henceforth, we refer to these paired halos as ``MW/M31 halos'', and we do not further differentiate between the MW and M31 given their similar masses \citep{VanDerMarel2012, BoylanKolchin2013}.
In the Appendix, we compare our results for paired versus isolated halos.

\subsection{Finding and tracking (sub)halos}
\label{sec:subhalo} 

ELVIS identifies dark-matter (sub)halos using the six-dimensional halo finder \textsc{rockstar} \citep{Behroozi2013a} and constructs merger trees using the \textsc{consistent-trees} algorithm \citep{Behroozi2013b}.
For each halo that is not a subhalo (see below), we assign a virial mass, $\mvir$, and radius, $\rvir$, using the evolution of the virial relation from \citet{BryanNorman1998} for our $\Lambda$CDM cosmology.
At $z = 0$, this corresponds to an overdensity of $\Delta_{\rm critical} = 97~(\Delta_{\rm matter} = 363)$ times the critical (matter) density, while at $z \gtrsim 3$ it asymptotes to $\Delta_{\rm critical} \approx \Delta_{\rm matter} \approx 178$.

We define a ``host halo'' as an isolated halo that can host (lower-mass) subhalos within in, and a ``subhalo'' as a halo whose center is inside $\rvir$ of a (more massive) host halo.
When a (sub)halo passes within $\rvir$ of a host halo, the (sub)halo becomes its ``satellite'' and experiences ``virial infall''.

For each (sub)halo, we assign its primary progenitor (main branch) as the progenitor that contains the most total mass summed from the (sub)halo masses over all preceding snapshots in that branch.
We then compute the peak mass, $\mpeak$, as the maximum instantaneous mass that a (sub)halo ever reaches along the history of its primary progenitor.
For subhalos, $\mpeak$ almost always occurs before virial infall.
As explored in \citet{GarrisonKimmel2014}, the numerical resolution of ELVIS does not significantly affect its (sub)halo catalogs at (and even below) $\mpeak > 10 ^ 8 \msun$, the limit in this work.

\subsection{Assigning stellar mass to subhalos}
\label{sec:stellar_mass} 

Our goal is to map luminous galaxies to the dark-matter subhalos in ELVIS.
The relation between stellar mass and subhalo mass (or maximum circular velocity) for dwarf galaxies is highly uncertain, likely with significant scatter, especially for our lowest-mass subhalos, some of which might not host any luminous galaxies, a manifestation of the long-standing ``missing satellites problem'' \citep{Klypin1999b}.
Nonetheless, we use the relation from abundance matching to ELVIS subhalos in \citet{GarrisonKimmel2014}, which is based on that of \citet{Behroozi2013c} but is modified at the low-mass end according to the observed stellar-mass function of \citet{Baldry2012}.
At the mass scales of dwarf galaxies, this leads to $\mstar \propto \mpeak ^ {1.92}$.
This modification reproduces the observed mass function at $\mstar < 10 ^ 9 \msun$ in the LG, especially if one accounts for observational incompleteness \citep{Tollerud2008, Hargis2014}.
However, we present most results as a function of both $\mstar$ and $\mpeak$, given the uncertainties of abundance matching at these low masses.
For reference, each of the MW/M31 halos hosts an average of 230, 28, and 3 satellites with $\mstar = 10 ^ {3 - 5}$, $10 ^ {5 - 7}$, and $10 ^ {7 - 9} \msun$, respectively, at $z = 0$ (this is similar for the paired and isolated halos).

In selecting the stellar-mass ratio for defining ``major'' groups or mergers in the histories of satellites in Sections~\ref{sec:preprocessing_v_mass} and \ref{sec:mergers}, we assume that the \textit{slope} (but not necessarily normalization) of this relation does not evolve, motivated by the lack of strong evolution observed for slightly more massive galaxies \citep[for example,][]{Leauthaud2012, Hudson2013} and the lack of observational evidence to suggest otherwise.
We define major mergers as those for which the ratio in $\mstar > 0.1$.
This broadly corresponds to mass ratios at which the lower-mass companion likely has significant dynamical effect on the more massive galaxy \citep[for example,][]{Hopkins2010, Helmi2012, YozinBekki2012} and for which recent mergers are likely to be observable.
Given our relation between $\mstar$ and $\mpeak$, this corresponds to a ratio in $\mpeak \gtrsim 0.3$.

\section{Virial-Infall Times of Satellites}
\label{sec:infall_time} 

We start by investigating the virial-infall times of satellite dwarf galaxies at $z = 0$, to understand how long they have been satellites within a host halo.
This timescale has several important implications.
First, it provides insight into how long satellites have experienced environmental processes that cause the observed depletion of gas, quenching of star formation, transition of morphology, and (potentially) stripping of stars.
Second, it tells us what fraction of satellites at $z = 0$ were satellites within a host halo during the epoch of cosmic reionization.
This allow us to understand whether the differing effects of reionization versus host-halo environment on surviving satellites occurred at distinct epochs in the formation histories of surviving satellites.

\subsection{Determining virial infall for satellites}
\label{sec:infall_time_definition} 

While environmental processes clearly affect satellite galaxies within $\rvir$ of the MW/M31 halos, whether such environmental processing occurs in lower-mass host halos ($\mvir \ll 10 ^ {12} \msun$) remains unclear.
Thus, we investigate two metrics of virial infall.
First, we examine ``first infall'': when a satellite first became a satellite within \textit{any} host halo, that is, first crossed within the virial radius of any halo more massive than itself.
We refer to this redshift as $\zinfallfirst$ or time as $\tinfallfirst$, with $\tsinceinfallfirst = t_{\rm now} - \tinfallfirst$.
We also examine ``MW/M31 infall'': when a satellite first became a satellite in its host MW/M31 halo.
We refer to this as $\zinfallhost$ or $\tinfallhost$, with $\tsinceinfallhost = t_{\rm now} - \tinfallhost$.

\subsection{Dependence of virial-infall time on satellite mass}
\label{sec:infall_time_v_mass}

\renewcommand{\figurewidth}{0.48}
\begin{figure*}
\centering
\includegraphics[width = \figurewidth \textwidth]{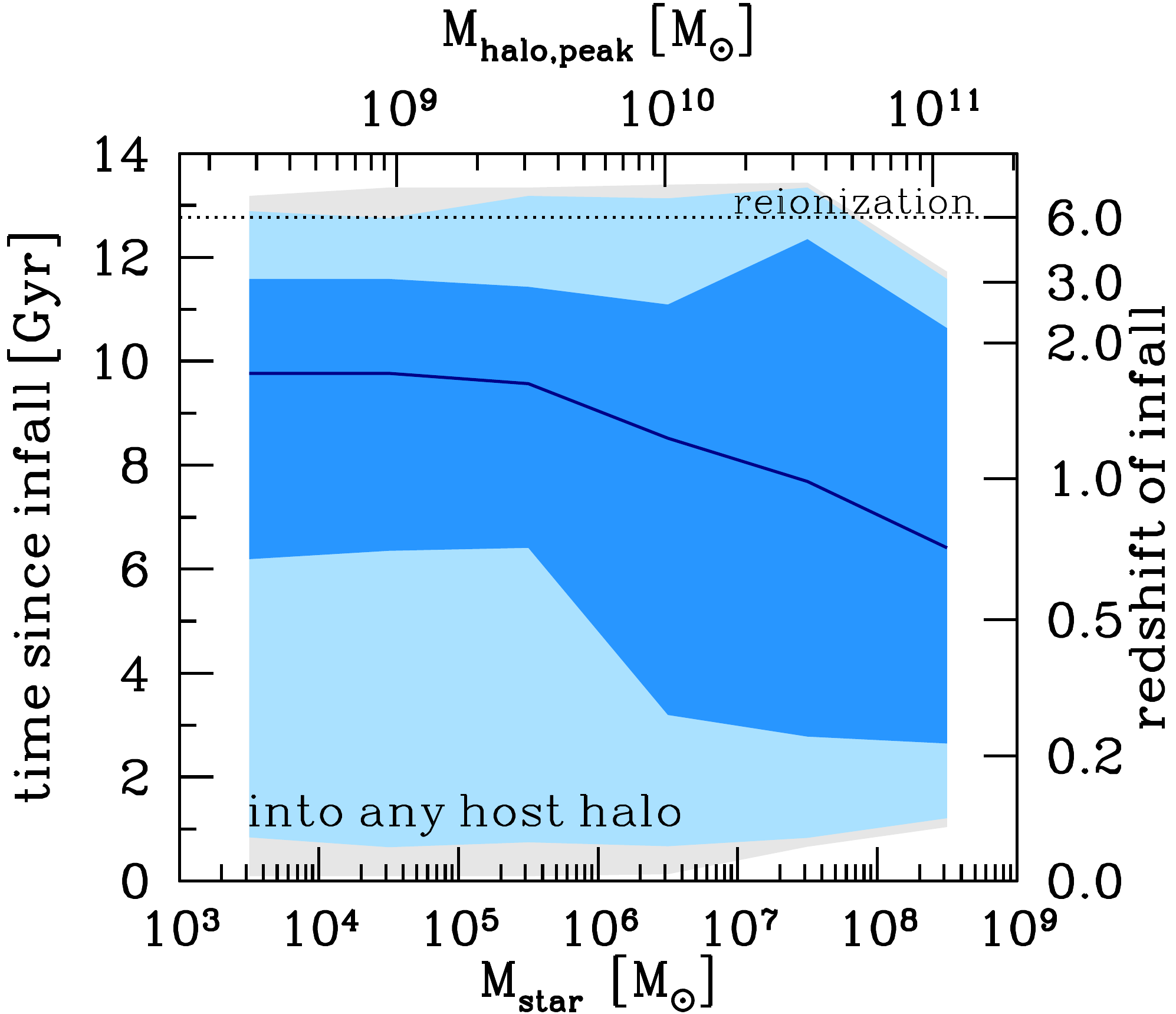}
\hspace{3 mm}
\includegraphics[width = \figurewidth \textwidth]{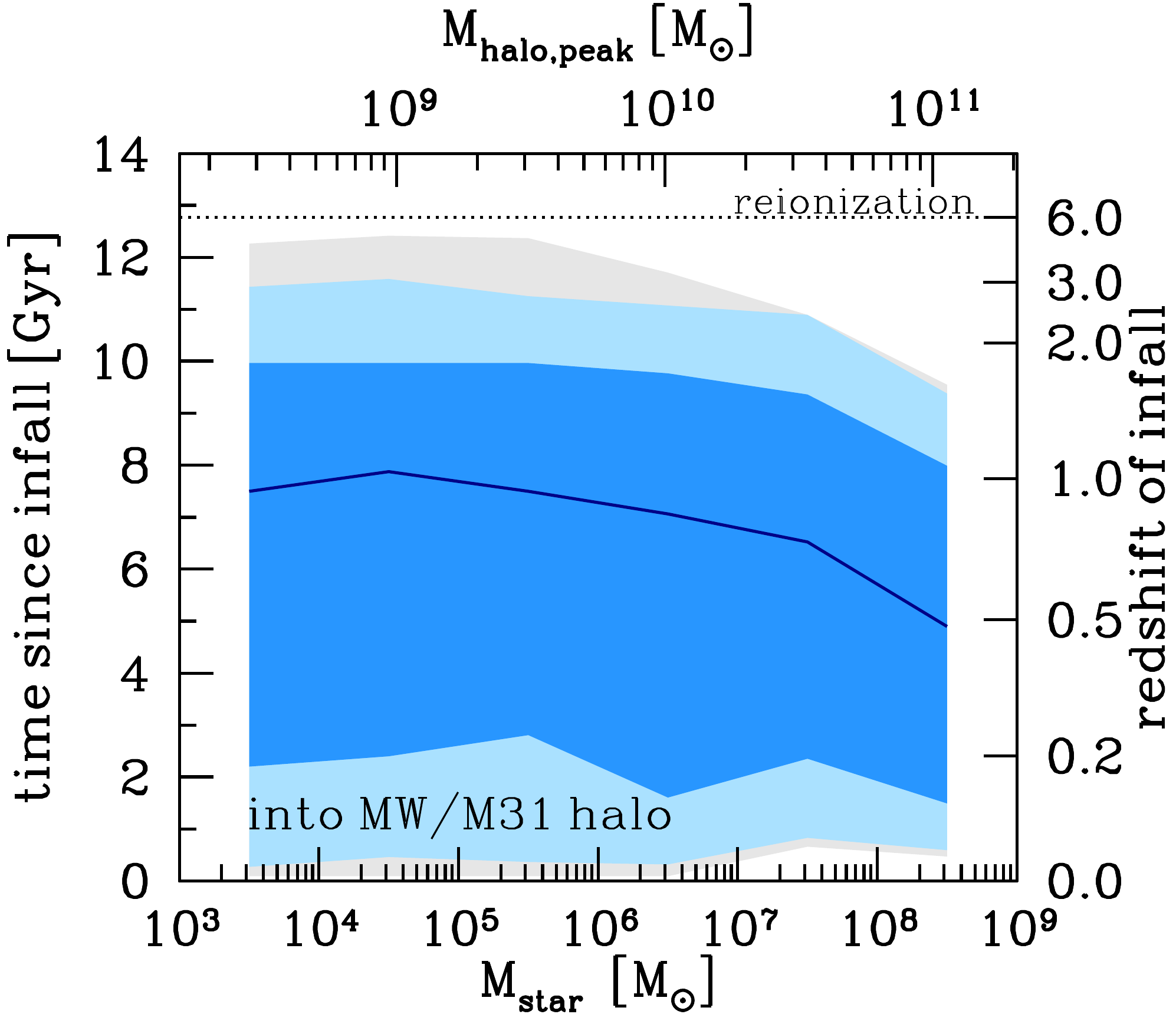}
\caption{
Time since virial infall for satellite dwarf galaxies at $z = 0$ as a function of their stellar mass, $\mstar$, or subhalo peak mass, $\mpeak$: time since first crossing within $\rvir$ of any host halo, thus including group preprocessing (left), or time since first crossing within $\rvir$ of the MW/M31 halo (right).
Curves show median, shaded regions show 68\%, 95\%, and 99.7\% of the distribution.
Satellites at $z = 0$ have been satellites typically for over half of their history.
Lower-mass satellites fell in earlier, though with large scatter.
Dotted line at $z = 6$ marks the end of cosmic reionization; during reionization, $< 4\%$ of current satellite galaxies were a satellite in a host halo, and \textit{none} were in the MW/M31 halo, demonstrating the effects of reionization versus host-halo environment occurred at distinct epochs, separated typically by $2 - 4 \gyr$, and thus are separable in time theoretically and, in principle, observationally, during satellites' evolutionary histories.
}
\label{fig:infall.time_v_mass}
\end{figure*}

We first examine how the virial-infall times of satellites at $z = 0$ depend on their mass.
Figure~\ref{fig:infall.time_v_mass} shows $\tsinceinfallfirst$ (left) and $\tsinceinfallhost$ (right), or $\zinfallfirst$ and $\zinfallhost$ on the right axes, as a function of satellite $\mstar$, or subhalo $\mpeak$ on the top axes.

For both virial-infall metrics, lower-mass satellites fell in earlier, though significant scatter persists at all masses.
This trend with $\mpeak$ (or $\mstar$) is a natural result of hierarchical structure formation, for two reasons.
First, halos of a given $\mpeak$ are more common at later cosmic time.
Thus, higher-mass satellites are more likely to have formed, and subsequently fallen in, at later time.
Second, satellites with higher $\mpeak$ have shorter dynamical-friction lifetimes (for fixed host-halo mass) before they tidally disrupt or merge with the host \citep{BoylanKolchin2008, Jiang2008, WetzelWhite2010}.

Our lowest-mass (ultra-faint) satellites first fell into any host halo typically $\sim 10 \gyr$ ago at $z \sim 1.7$, and they first fell into the MW/M31 halo $\sim 7.5 \gyr$ ago at $z \sim 1$.
By contrast, our highest-mass satellites (corresponding to the LMC, SMC, NGC 205, M32) have $\tsinceinfallfirst \sim 6.5 \gyr$ ($\zinfallfirst \sim 0.8$) and $\tsinceinfallhost \sim 5 \gyr$ ($\zinfallhost \sim 0.5$).
Thus, satellites first fell into any host halo a few Gyr before they fell into the MW/M31 halo, as we will examine further in Section~\ref{sec:preprocessing_duration_mass}.
This result is \textit{generic} for hierarchical structure formation.

Overall, satellite dwarf galaxies at $z = 0$ typically have evolved as a satellite in a host halo for over half of their entire history, so the host-halo environment typically has had significant time to affect their evolution.

\subsection{Dependence of virial-infall time on satellite distance}
\label{sec:infall_time_v_distance}

\renewcommand{\figurewidth}{0.33}
\begin{figure*}
\centering
\includegraphics[width = \figurewidth \textwidth]{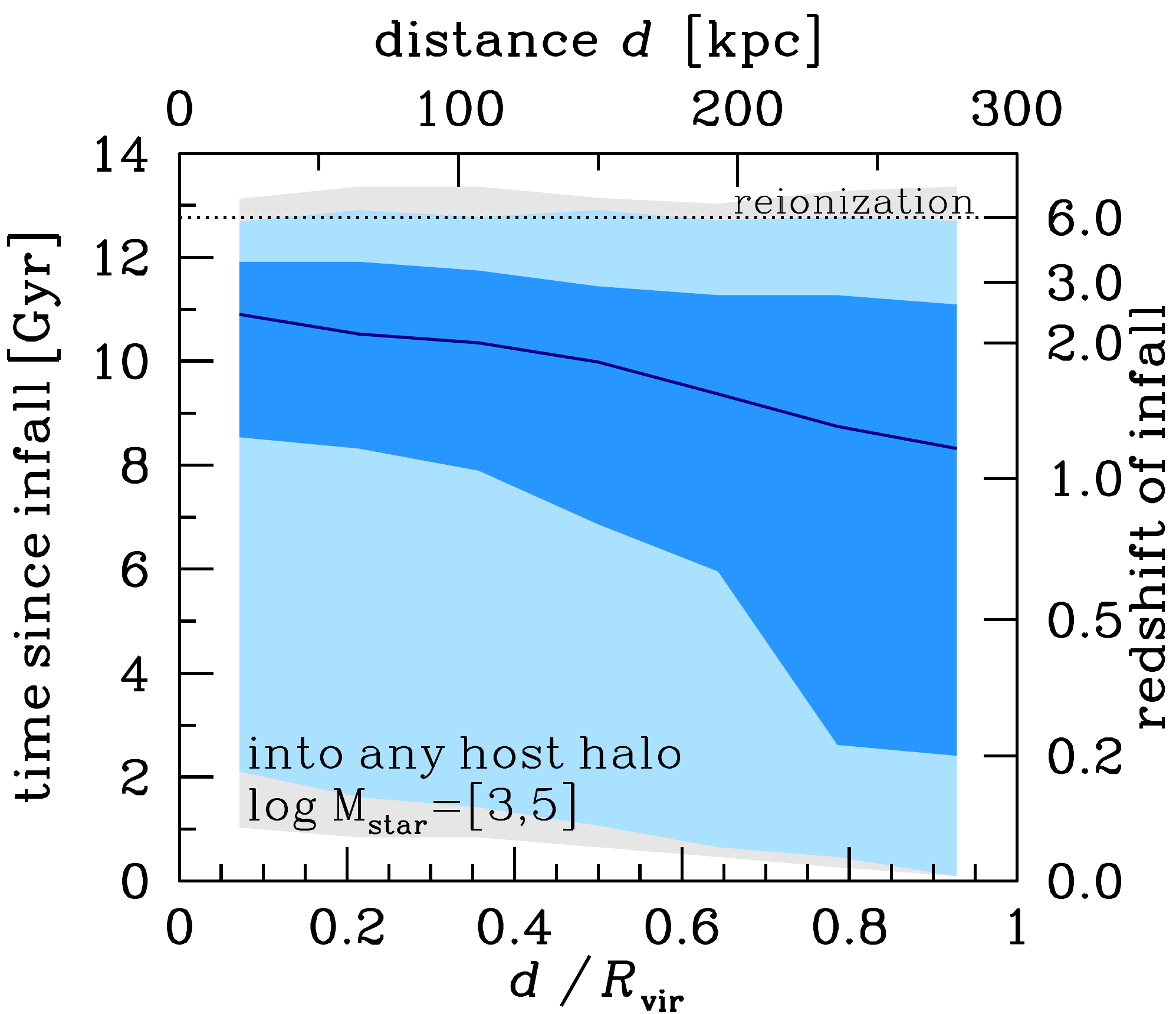}
\includegraphics[width = \figurewidth \textwidth]{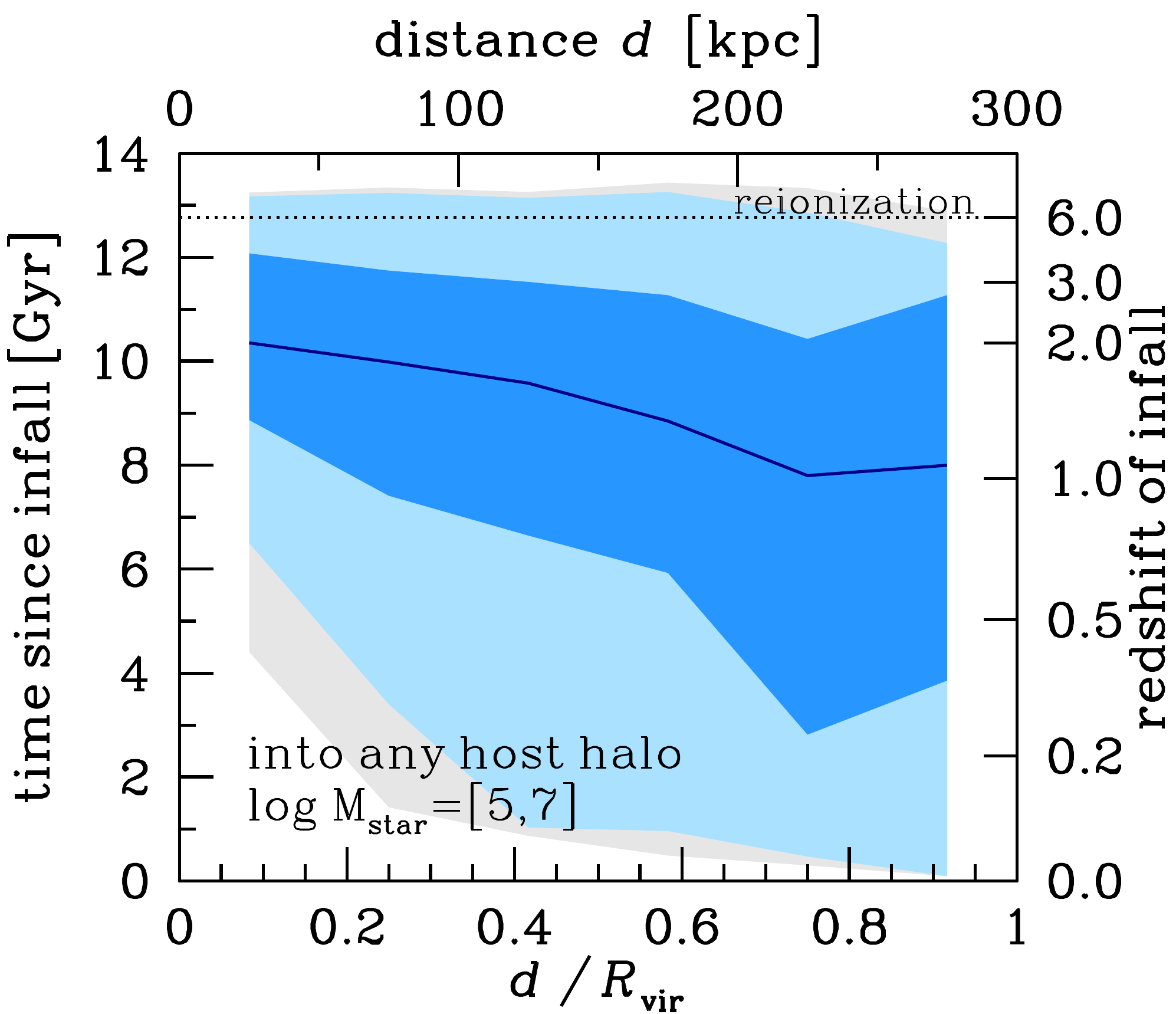}
\includegraphics[width = \figurewidth \textwidth]{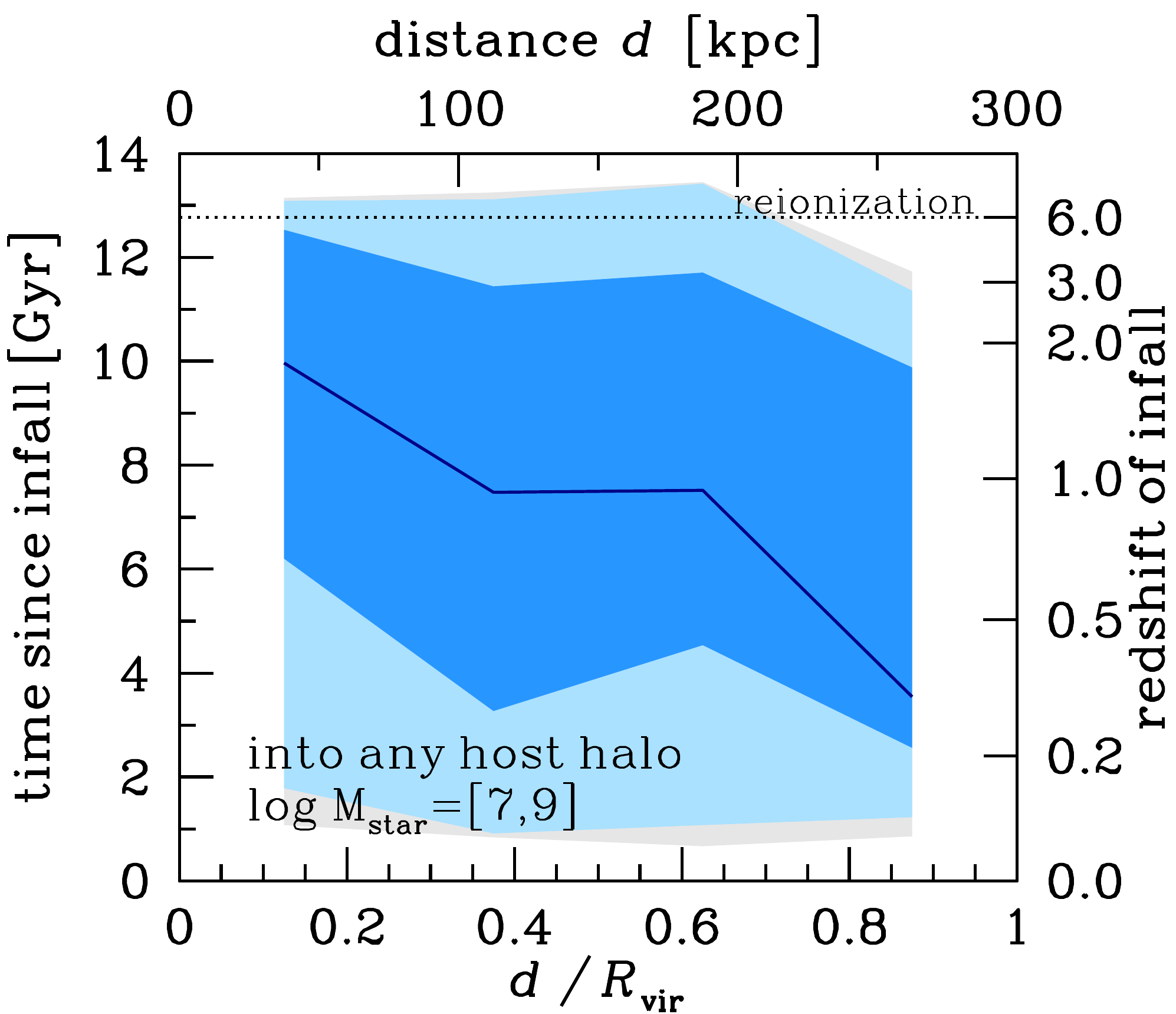}
\includegraphics[width = \figurewidth \textwidth]{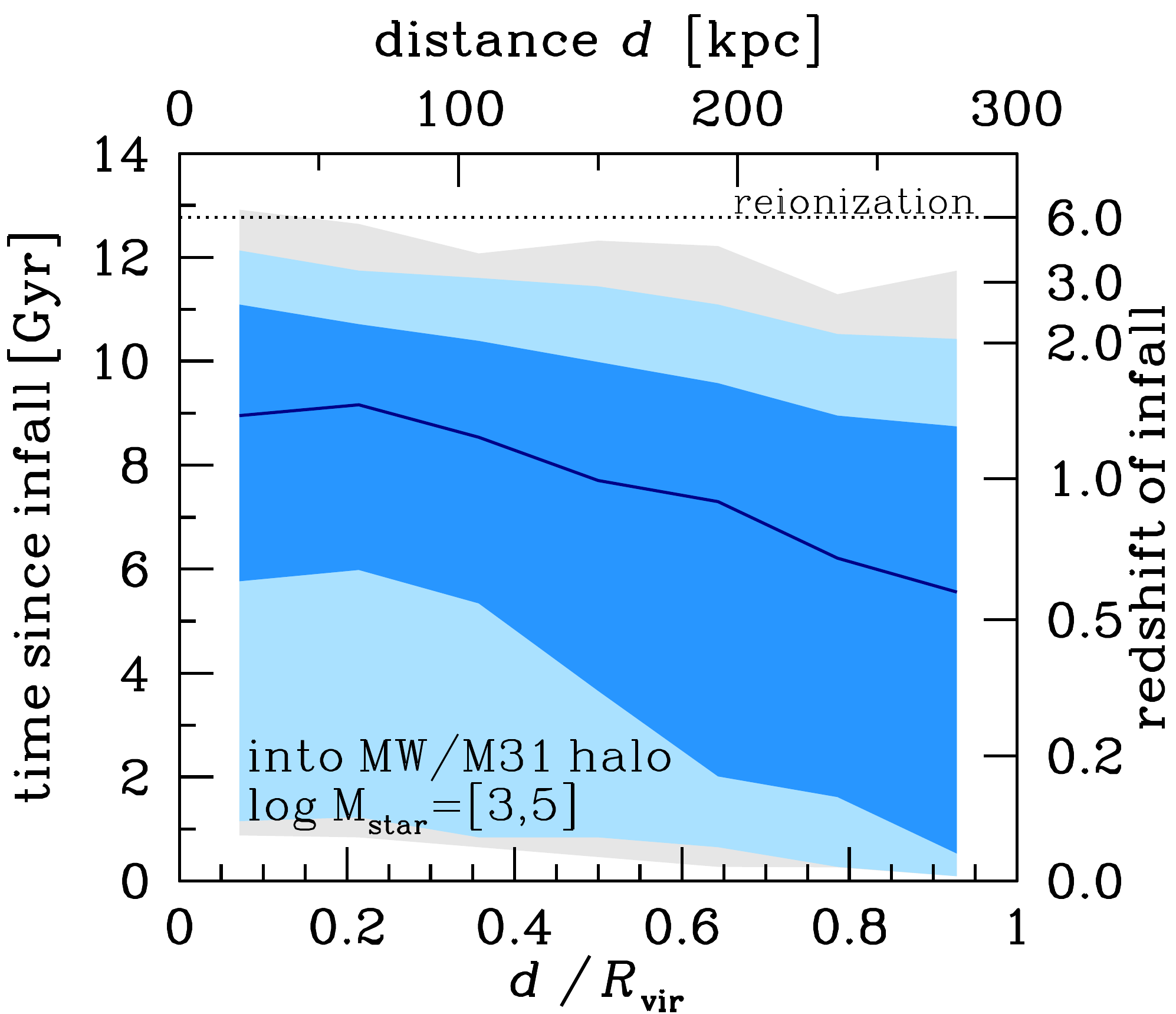}
\includegraphics[width = \figurewidth \textwidth]{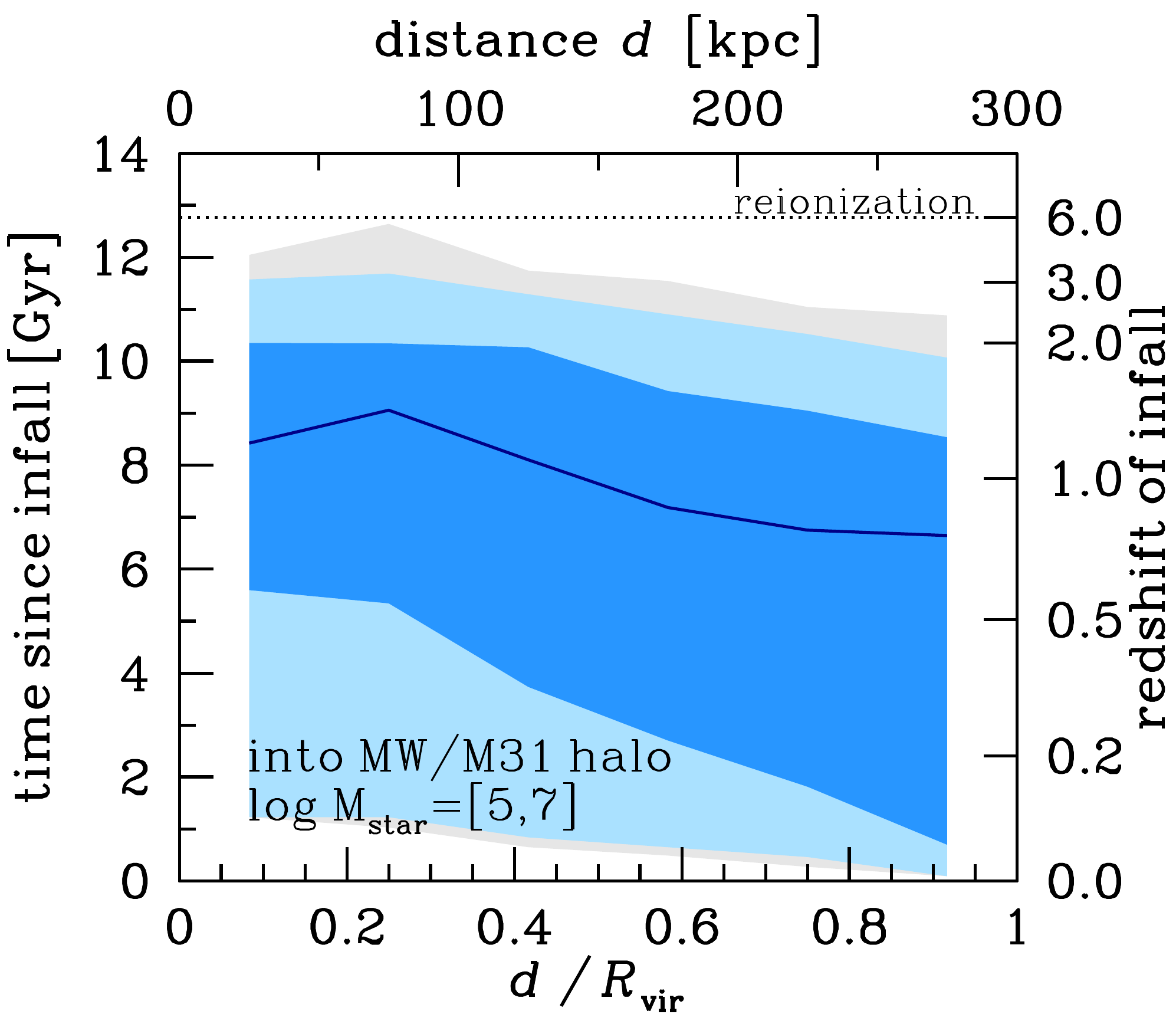}
\includegraphics[width = \figurewidth \textwidth]{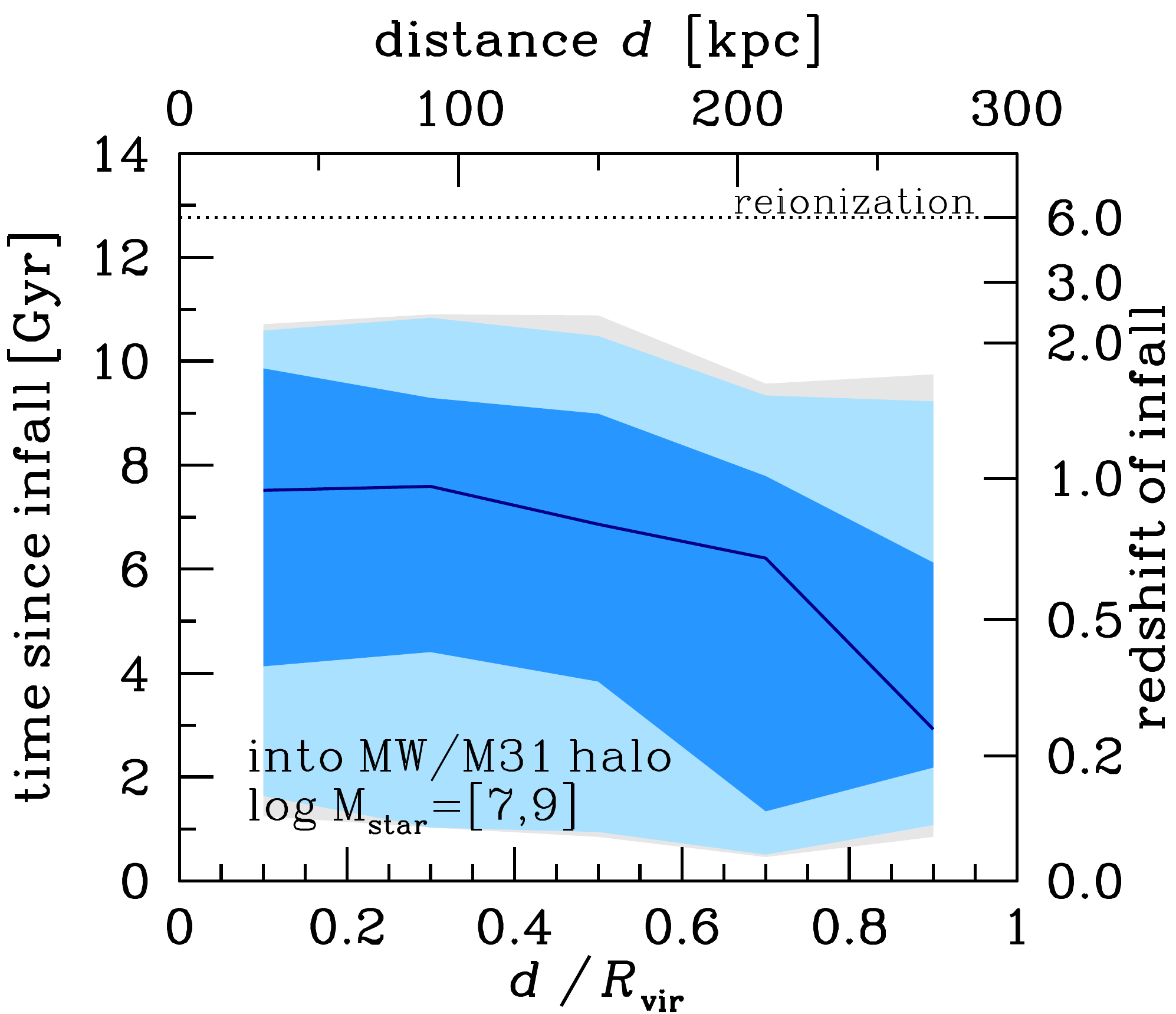}
\caption{
Time since virial infall for satellite dwarf galaxies at $z = 0$ as a function of their distance to their host, $d$, or as scaled to the host's virial radius, $\rvir$, at $z = 0$: time since first crossing within $\rvir(z)$ of any host halo (top row) or within $\rvir(z)$ of the MW/M31 halo (bottom row).
Left-to-right columns show satellites of increasing $\mstar$.
Curves shows the median, shaded regions show 68\%, 95\%, and 99.7\% of the distribution.
At all masses, satellites closer to the host fell in earlier, though with significant scatter.
Dotted line at $z = 6$ indicates the end of cosmic reionization; even for the lowest-mass (ultra-faint) satellites at the smallest distances, where they are observable, $< 4\%$ were a satellite in any host halo during reionization, and \textit{none} were in the MW/M31 halo during reionization.
}
\label{fig:infall.time_v_distance}
\end{figure*}

We next explore how the above virial-infall times of satellites depend on their current distance from the MW/M31 host.
Previous analyses at higher mass scales in cosmological simulations showed that satellites at smaller distances tend to have fallen in earlier \citep[for example,][]{Gao2004b, Smith2012, Oman2013}.
This is because (1) $\rvir$ of the host halo was smaller at earlier time, and (2) satellite orbits get dragged to smaller distance over time via dynamical friction.
Such a trend for satellites in the LG would be important for several reasons.
First, ultra-faint ($\mstar \lesssim 10 ^ 5 \msun$) satellites currently are observable only within the inner $\sim 50 \kpc$ of the MW halo, so it is possible that they had systematically earlier infall times than the median in Figure~\ref{fig:infall.time_v_mass}.
Second, such a correlation with distance would provide a statistical proxy for an environmental evolutionary sequence (since the time of infall) for the observable satellite population.

Figure~\ref{fig:infall.time_v_distance} shows $\tsinceinfallfirst$ (top row) and $\tsinceinfallhost$ (bottom row) as a function of distance to the host's center, $d$, as scaled to the host's virial radius, $\rvir$, at $z = 0$.
We compute these quantities in bins of $d / \rvir$ for each MW/M31 halo, and using that the median $\rvir$ in ELVIS is $300 \kpc$, we also show the (approximate) dependence on $d$ along the top axes.
For both infall metrics and at all masses, satellites closer to the host fell in earlier, though with significant scatter.
Overall, the trend for MW/M31 infall is slightly stronger than for first infall, as expected given that we measure $d$ with respect to the MW/M31 center.
The gradient in $\tsinceinfallfirst$ and $\tsinceinfallhost$ from $d / \rvir = 0$ to 1 for massive satellites is $6.5$ and $4.5 \gyr$, respectively, while for our lowest-mass satellites it is $2.5$ and $4 \gyr$.
Thus, the correlation of infall time with distance is stronger for more massive satellites, because (1) they fell in more recently, making them less smeared out in orbital phase space, and (2) they experience more efficient dynamical friction.
For our lowest-mass (ultra-faint) satellites, those in our smallest distance bin, where they are observable, experienced first infall typically $\sim 11 \gyr$ ago at $z \sim 2.2$, and they first fell into the MW/M31 halo $\sim 9 \gyr$ ago at $z \sim 1.5$, so these are slightly earlier than Figure~\ref{fig:infall.time_v_mass}.

We conclude that a satellite's distance does provide a statistical proxy for its virial-infall time, and therefore, the distribution of distances for the satellite population at $z = 0$ provides a proxy for an environmental evolutionary sequence, especially for more massive satellites.
However, note one caveat: despite the correlation with distance, satellites currently near $\rvir$ have been satellites for quite a while, typically $3 - 8 \gyr$, depending on mass.
Thus, the satellite population near $\rvir$ is not \textit{only} a recent-infall population, but rather, it is a superposition of inward- and outward-orbiting satellites, some of which already experienced a pericentric passage.
Note that any given satellite spends most of its orbital time near apocenter.
(For reference, the virial crossing time, $\rvir / \vvir$, is $\approx 2 \gyr$ at $z = 0$.)

\subsection{Implications for dwarf galaxies during cosmic reionization}
\label{sec:reionization}

The virial-infall times in Figures~\ref{fig:infall.time_v_mass} and \ref{fig:infall.time_v_distance} have important implications for understanding the relative effects of cosmic reionization versus host-halo environment on surviving dwarf galaxies, in particular, whether the effects of these two processes occurred at distinct epochs during the formation histories of these galaxies.
In these figures, the dashed line at $z = 6$ represents the end of cosmic reionization as constrained by various observations \citep[for example,][and references therein]{Robertson2013}.
Across all masses and distances, \textit{none} of the satellites at $z = 0$ were within $\rvir$ of their MW/M31 halo any time during reionization.
Furthermore, $< 4\%$ were within $\rvir$ of \textit{any} host halo during reionization.
There are regimes where this fraction is somewhat higher, the highest being 10\% for satellites at $\mstar = 10 ^ {5 - 8} \msun$ and $d / \rvir < 0.1$, but this fraction is typically only a few percent across our range of mass and distance.
Thus, essentially none of the dwarf galaxies in the LG were within $\rvir$ of a host halo during reionization, such that they would have experienced strong environmental effects at that time.

To understand this result in more detail, we also examine how close the dwarf galaxies came to a more massive halo during reionization.
Thus, we select all satellites in the MW/M31 halos at $z = 0$ and trace them back to $z > 6$, when almost all such dwarfs were isolated (non-satellite) halos.
(We are able to track all (sub)halos back to $z > 6$, except for 6\% of those at $\mstar < 10 ^ 4 \msun$, which formed after $z = 6$.)
At all $z > 6$ (for which ELVIS contains 6 snapshots), we then compute the nearest distance, $d_{\rm nearest}$, that each satellite's progenitor came to the center of any neighboring halo that is more massive and thus feasibly could induce environmental effects.

Figure~\ref{fig:nearest_distance_reionization} shows the cumulative distribution of $d_{\rm nearest}$ in comoving units (left) and scaled to $\rvir$ of the nearest, more massive halo (right).
The thick solid curve shows the average across the paired MW/M31 halos.
We show the average for all satellites across our mass range, $\mstar = 10 ^ {3 - 9} \msun$, as we find no significant dependence on satellite mass.
The typical $d_{\rm nearest}$ at $z > 6$ was large: $3 \mpc$ comoving ($\sim 400 \kpc$ physical), or $500 \, \rvir$.
Moreover, only $\approx 5\%$ of these galaxies came within $1 \mpc$ comoving, or $100 \, \rvir$, of a more massive halo.
The thin solid curves show each MW/M31 pair, highlighting the factor of $\sim 2$ scatter in these distributions.
For reference, the thick dotted curve shows the average for the isolated MW/M31 halos, which we discuss in the Appendix.

If strong environmental effects (such as star-formation quenching) on dwarf galaxies are confined to within $\rvir$ (or even larger) of a host halo at these redshifts, then the results of Figures~\ref{fig:infall.time_v_mass}, \ref{fig:infall.time_v_distance}, and \ref{fig:nearest_distance_reionization} indicate that such environmental processing occurred only at $z < 6$ during the histories of satellites at $z = 0$, and more typically, at $z \lesssim 3$ (for $\approx 84\%$ of all satellites).
The properties of the LG support this result, because a strong transition in morphology, star formation, and gas content of dwarf galaxies, as induced by the host-halo, occurs only within $\approx 300 \kpc$ ($\approx \rvir$) of the MW or M31.

Given that cosmic reionization ended by $z = 6$, we thus conclude that the effects of reionization occurred well before those of the host-halo environment for surviving satellites.
Specifically, for surviving satellites, the duration between the end of reionization and first crossing within $\rvir$ of a host halo was typically $> 2 \gyr$ for first infall into any host halo (including group preprocessing) and $> 4 \gyr$ for first infall into the MW/M31 halo.
This result strongly motivates the use of faint and ultra-faint satellites as probes of reionization, for example, by measuring the (potential) effect of reionization on their star-formation histories as derived from their current stellar populations \citep[for example,][]{Brown2014, Weisz2014b}, if such star-formation histories can demonstrate that any individual satellite quenched at $z \gtrsim 6$ or that any common features (such as quenching) occurred at $z \gtrsim 3$ for a population of several satellites (given that $\approx 84\%$ surviving satellites fell into a host halo at $z \lesssim 3$).

\renewcommand{\figureheight}{0.30}
\begin{figure*}[h!]
\centering
\includegraphics[height = \figureheight \textheight]{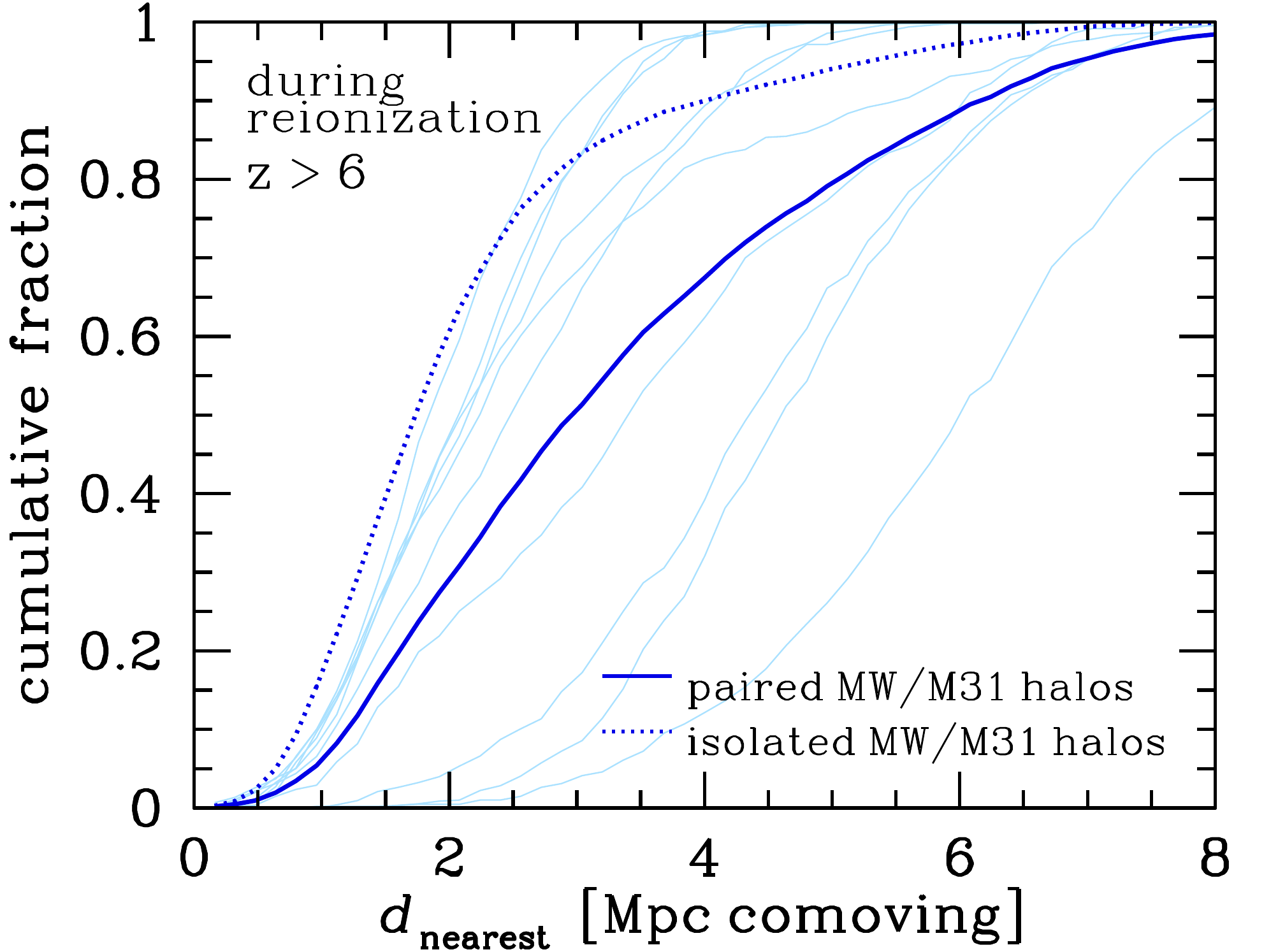}
\includegraphics[height = \figureheight \textheight]{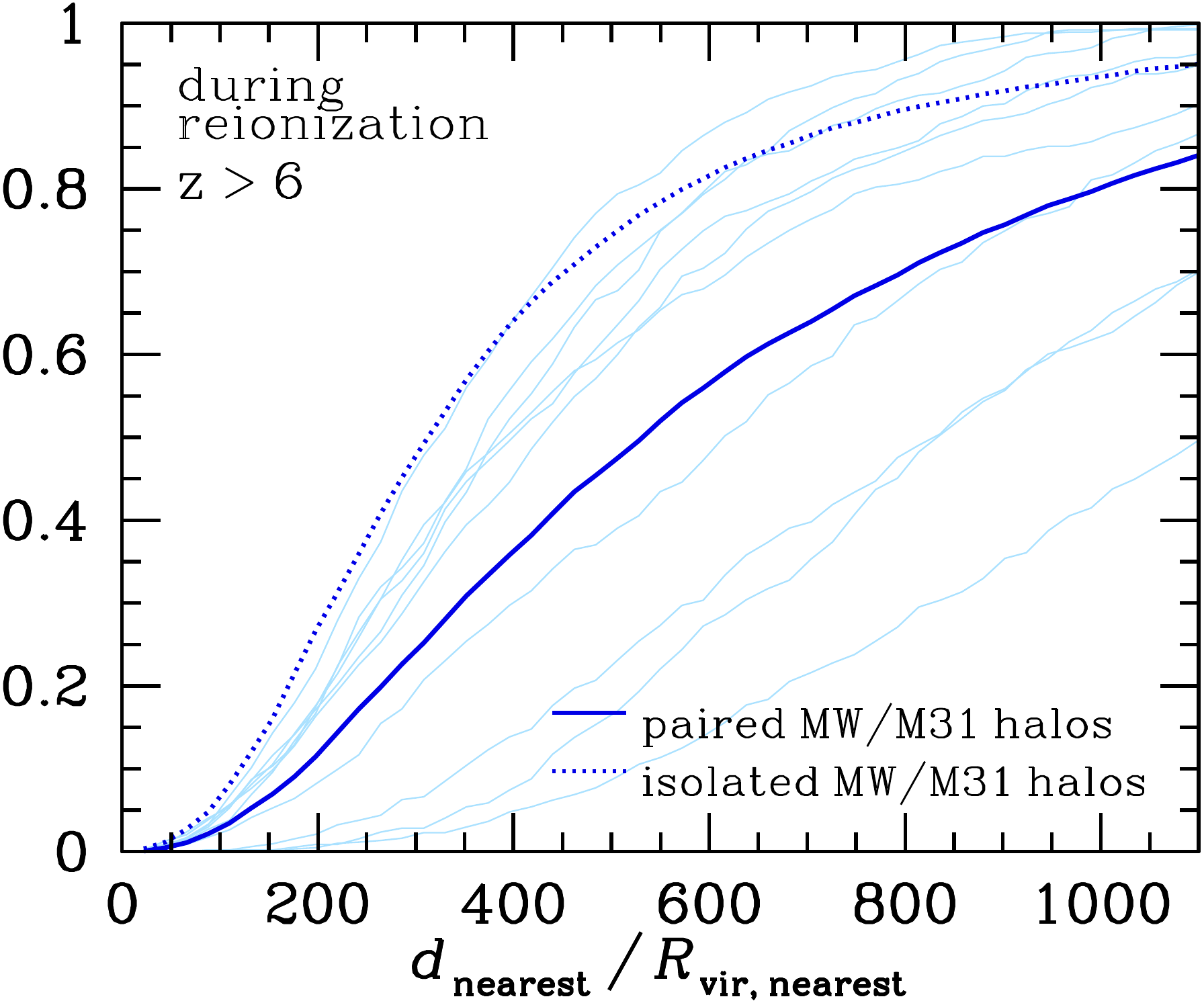}
\caption{
For all satellite galaxies at $z = 0$ with $\mstar = 10 ^ {3 - 9} \msun$, cumulative distribution of the distance to the nearest, more massive halo, $d_{\rm nearest}$, that they experienced during cosmic reionization ($z > 6$).
Left panel shows comoving distance, and right panel shows this distance scaled to $\rvir$ of the nearest halo.
Solid thick curve shows average over all satellites in the paired MW/M31 halos, while thin curves shows satellites in each pair, to indicate the pair-to-pair scatter.
We find no dependence on satellite mass.
The typical distance was $3 \mpc$ comoving ($\sim 400 \kpc$ physical), or $500 \, \rvir$; at these distances, dwarf galaxies in the Local Group do not show strong environmental influence.
These results strongly support that the effects of the host-halo environment occurred well after the effects of reionization during the histories of surviving satellites.
For comparison, dotted curve shows the average across the isolated MW/M31 halos, whose satellites experienced $\sim 1/2$ the distance (see Appendix).
}
\label{fig:nearest_distance_reionization}
\end{figure*}

\section{Group Infall and Preprocessing}
\label{sec:preprocessing} 

In the previous section, we showed that many satellite galaxies first became satellites significantly prior to falling into the MW/M31 halo.
Thus, many satellites spent significant time in another host halo, which may have environmentally ``preprocessed'' them prior to their joining the MW/M31 halo.
We now explore what fraction of all satellites at $z = 0$ were preprocessed as a satellite in another host halo (group) prior to falling into the MW/M31 halo.
We examine two such metrics: the fraction of all current satellites that were a satellite in another host halo (1) \textit{any time before} falling into the MW/M31 halo, or (2) \textit{at the time of} falling into the MW/M31 halo.
The difference between these is driven by ``ejected'' or ``backsplash'' satellites that fell into another host halo and then orbited out beyond its $\rvir$ before falling inside $\rvir$ of the MW/M31 halo.
Recent work indicates that these satellites are affected environmentally in similar ways as satellites that remain within $\rvir$ \citep[for example,][]{Ludlow2009, Knebe2011, Teyssier2012, Bahe2013, Wetzel2014}.
Thus, we examine both preprocessing metrics.

\subsection{Dependence of group preprocessing on satellite mass}
\label{sec:preprocessing_v_mass}

\renewcommand{\figureheight}{0.33}
\begin{figure*}
\centering
\includegraphics[height = \figureheight \textheight]{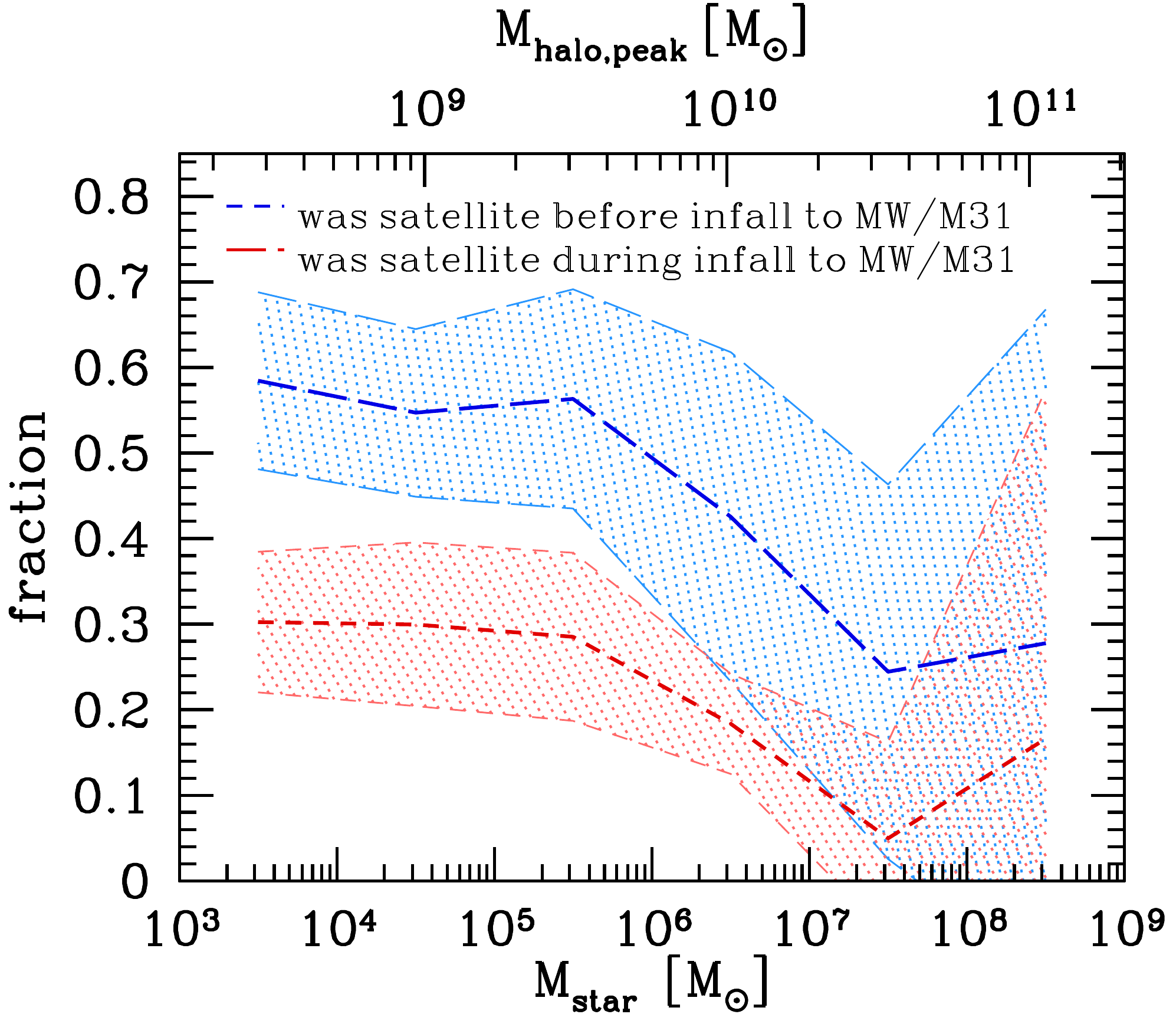}
\includegraphics[height = \figureheight \textheight]{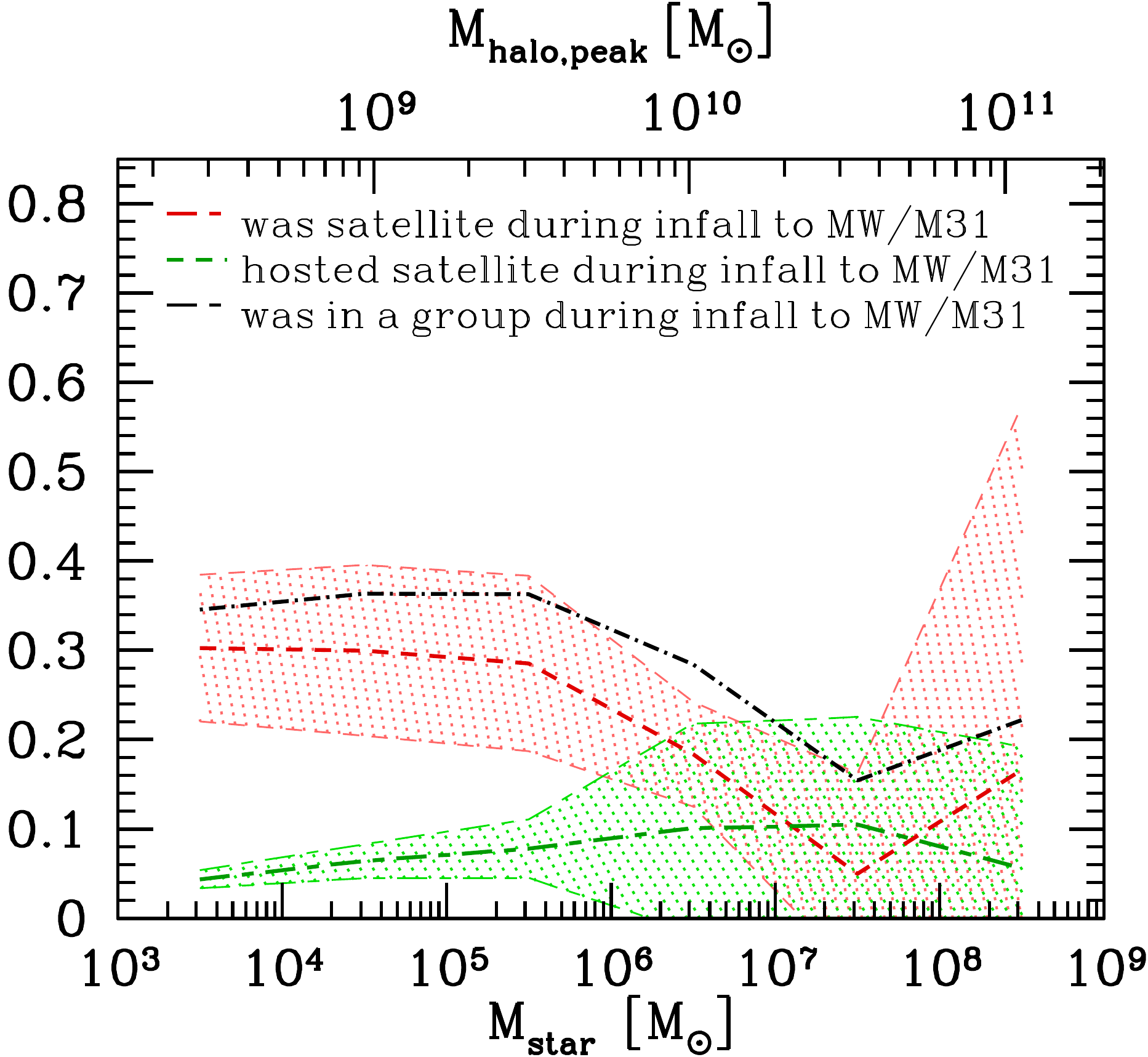}
\caption{
Fraction of all satellite dwarf galaxies at $z = 0$ that experienced various aspects of group preprocessing prior to falling into the current MW/M31 halo, as a function of satellite stellar mass, $\mstar$, or subhalo peak mass, $\mpeak$.
\textbf{Left}: Fraction that were a satellite in a group \textit{any time prior to} (blue long-dashed) or \textit{at the time of} (red short-dashed) falling into the MW/M31 halo.
Curves show average over the paired MW/M31 halos, while shaded regions show standard deviation from halo-to-halo scatter.
The difference between the two curves is driven by ejected/backsplash satellites that were once within another host halo but then orbited beyond its $\rvir$.
Lower-mass satellites are more likely to have been preprocessed in a group: over \textit{half} of the lowest-mass (faint and ultra-faint) satellites were.
\textbf{Right}: Red short-dashed curve shows same as in the left panel, while green long-short-dashed curve shows the fraction that were the central (most massive) galaxy in a group that contained a major satellite (at least $0.1 \times$ its $\mstar$) at the time of falling into the MW/M31 halo.
Black dot-dashed curve show the sum of the two curves.
}
\label{fig:infall.fraction_v_mass}
\end{figure*}

Figure~\ref{fig:infall.fraction_v_mass} (left) shows both of the above preprocessed fractions as a function of satellite $\mstar$ or $\mpeak$.
First, the red short-dashed curve shows the fraction of satellites that were a satellite in a group at the time of falling into the MW/M31 halo.
For our highest-mass satellites, this fraction is relatively low ($\sim 10\%$) but increases significantly to 30\% for our lowest-mass satellites.
Thus, $\sim 1 / 3$ of all faint and ultra-faint satellites were in a group when they fell into the MW/M31 halo, in good agreement with the results of \citet{LiHelmi2008}, which was based on a cosmological simulation of a single MW-like halo.

Second, the blue long-dashed curve shows the fraction of satellites that were a satellite in a group \textit{any time} before falling into the MW/M31 halo.
This fraction is significantly ($\sim 2 \times$) higher, because of the large fraction of satellites whose orbits brought them beyond $\rvir$ of their preprocessing group.
For our highest-mass satellites, $\sim 30\%$ were preprocessed by a group before joining the MW/M31 halo.
Again, this preprocessed fraction increases at lower mass, being $\sim 60\%$ for faint and ultra-faint satellites.
Thus, \textit{half} of all satellite galaxies with $\mstar < 10 ^ 6 \msun$ were preprocessed as a satellite in a group prior to joining the MW/M31 halo.

We also explore how many current satellites in the MW/M31 halo were the central (most massive) galaxy in such an infalling group.
That is, we identify satellites in the MW/M31 halo at $z = 0$ that hosted their own major satellite(s) when they fell into the MW/M31 halo.
(By ``major'', we mean that $\mstar$ differs by less than a factor of 10, as detailed in Section~\ref{sec:stellar_mass}).
Figure~\ref{fig:infall.fraction_v_mass} (right) shows this fraction via the green long-short-dashed curve, which is $5 - 10\%$ across our mass range.
This fraction increases only weakly with mass because of the combination of (1) a nearly mass-independent distribution of $M_{\rm peak,\,satellite} / M_{\rm peak,\,host}$ in $\Lambda$CDM, (2) our power-law $\mstar - \mpeak$ relation, and (3) our requirement that $M_{\rm star,\,satellite} / M_{\rm star,\,host} > 0.1$.

For comparison, Figure~\ref{fig:infall.fraction_v_mass} (right) also shows the same red short-dashed curve as in the left panel and the black dot-dashed curve shows the sum of the two curves, indicating the total fraction of satellites that fell into the MW/M31 halo as part of a major group.
This overall fraction is significant at $20 - 40\%$ across our mass range.

\subsection{Dependence of group preprocessing on satellite distance}
\label{sec:preprocessing_v_distance}

\renewcommand{\figureheight}{0.33}
\begin{figure*}
\centering
\includegraphics[height = \figureheight \textheight]{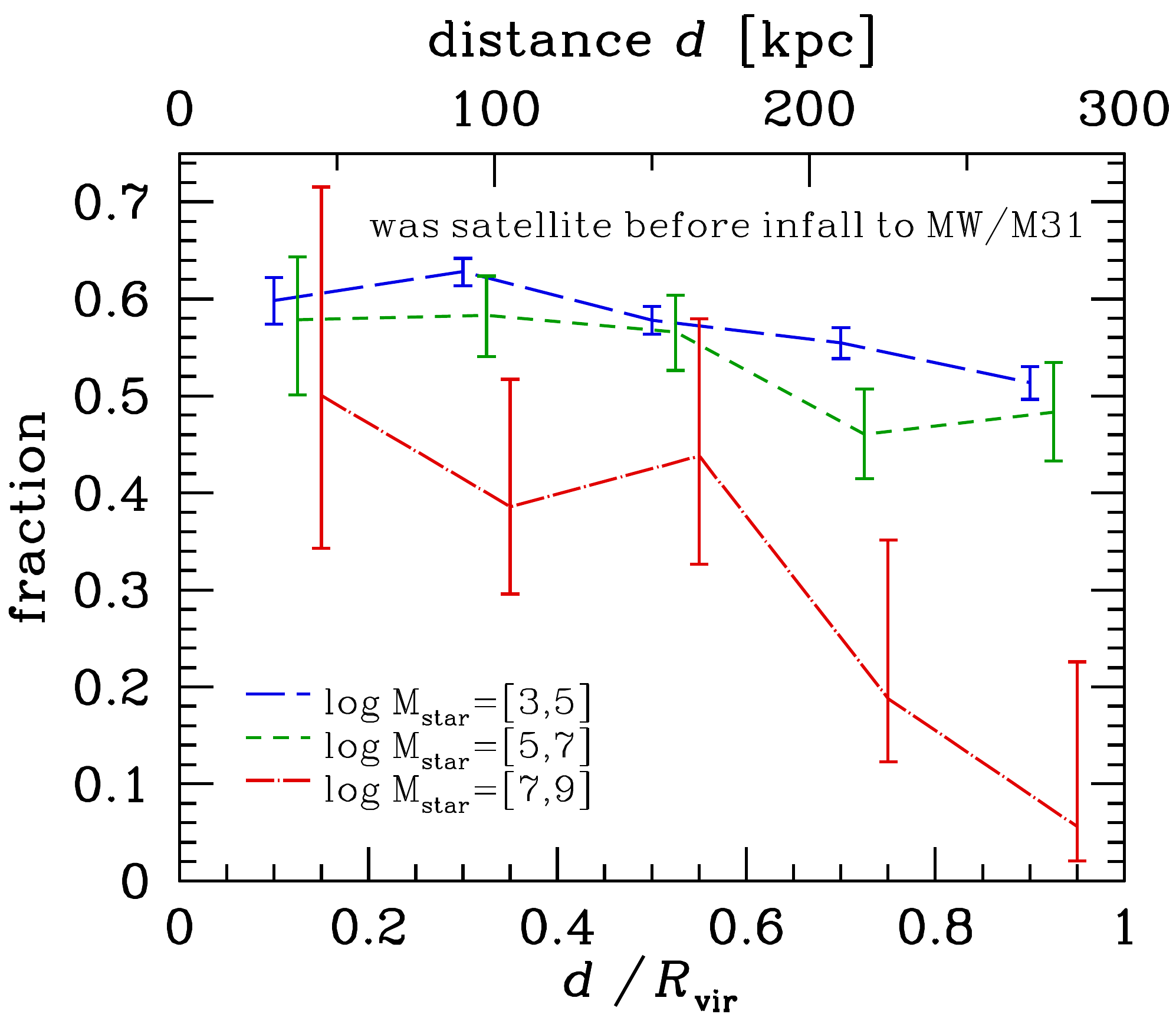}
\includegraphics[height = \figureheight \textheight]{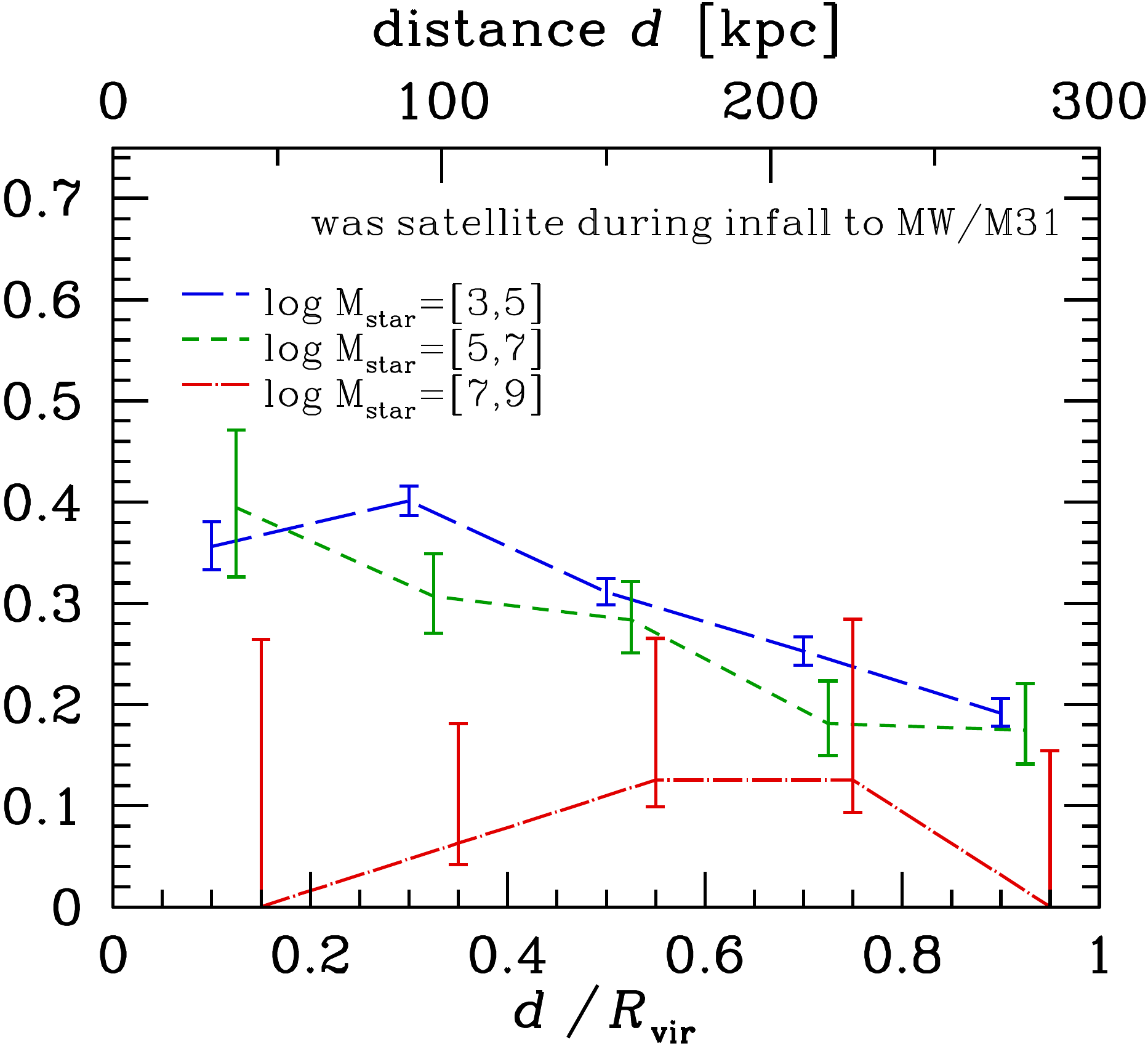}
\caption{
Fraction of all satellite dwarf galaxies at $z = 0$ that were a satellite in another host halo \textit{any time prior to} (left) or \textit{at the time of} (right) falling into the MW/M31 halo, as a function of their current distance to the host, $d$, or as scaled to the host's virial radius, $\rvir$.
Curves show average over the paired MW/M31 halos in bins of satellite $\mstar$, and error bars show 68\% uncertainty in this fraction for a beta distribution (the halo-to-halo scatter is similar to Figure~\ref{fig:infall.fraction_v_mass}).
For both group-preprocessing metrics, and at across all masses, satelltes closer to the host are more likely to have been preprocessed.
}
\label{fig:infall.fraction_v_distance}
\end{figure*}

We next explore how the above preprocessed fractions vary with the current distance of satellites from their host.
The two panels in Figure~\ref{fig:infall.fraction_v_distance} shows the same two preprocessed fractions as in Figure~\ref{fig:infall.fraction_v_mass} (left), but as a function of $d / \rvir$ at $z = 0$, similar to Figure~\ref{fig:infall.time_v_distance}.
Again, we compute these quantities in bins of $d / \rvir$ for each MW/M31 halo, though we also show the dependence on $d$ (using the median $\rvir$ across the MW/M31 halos) along the top axes.
At any distance, lower-mass satellites are more likely to have been preprocessed.
Moreover, at nearly all satellite masses, those closer to the host center are more likely to have been preprocessed, with a nearly $2 \times$ increase from $d / \rvir = 1$ to 0.1 for our lowest-mass satellites.
Most likely, this gradient arises because a satellite that fell in as part of a group remained bound to the (more massive) group for some time after infall, so it experienced more efficient dynamical friction that dragged it to the center of the MW/M31 halo more rapidly.

Overall, given that the faintest galaxies are observable only at small distances within the MW halo, most likely about \textit{half} of them were a satellite in another group(s) before / during infall into the MW halo.

\subsection{Duration and host-halo mass of group preprocessing}
\label{sec:preprocessing_duration_mass} 

We next examine in more detail the durations and host-halo masses that satellites at $z = 0$ experienced during their group preprocessing, in order to understand better its potential impact on their evolution.

Figure~\ref{fig:preprocessing_time_mass_v_mass} (top) shows the distribution of maximum host-halo mass that satellites experienced during group preprocessing as a function of their $\mstar$.
The typical preprocessing group had $\mvir \sim 10 ^ {11} \msun$, with 68\% spread of $10 ^ {10 - 12} \msun$, largely independent of satellite mass, though with scatter to $\mvir$ at lower $\mstar$.
Combined with Figure~\ref{fig:infall.fraction_v_mass} (left), this means that $\sim 25 \%$ of all satellites at $z = 0$ were preprocessed in groups of $\mvir \gtrsim 10 ^ {11} \msun$, comparable to the LMC or M33.

The total population of preprocessed satellites at $z = 0$ originated from typically 30 infalling groups, with most groups containing $2 - 5$ satellites.
However, about half of all preprocessed satellites fell in via the $2 - 3$ most massive groups ($\mvir \gtrsim 10 ^ {11} \msun$, similar to the LMC or M33), and the most massive infalling group typically brought in $\sim 25 - 30 \%$ of all preprocessed satellites, or $\sim 13 \%$ of all satellites in total.

However, for half of the preprocessed satellites, their preprocessing host does not survive to $z = 0$ but instead merges/disrupts within the MW/M31 halo.
Thus, surviving preprocessed satellites do not necessarily have their preprocessing host galaxy 
surviving near them at $z = 0$; instead, the host galaxy could persist in a disrupted stream configuration.
On the other hand, particularly massive ($\mpeak \gtrsim 10 ^ {11} \msun$, comparable to the LMC) satellites that do survive to $z = 0$ typically brought in $\sim 7 \%$ of overall satellite dwarf population in the MW/M31 halo \citep{Deason2015}.

Satellites can be preprocessed by the other MW/M31 halo in the pair, if they fell into one of the MW/M31 halos, orbited beyond its $\rvir$, and then fell into the other MW/M31 halo.
However, this accounts for $< 2\%$ of all preprocessed satellites across our mass range, in general agreement with \citet{Knebe2011}, so this is not a particularly important population.

Figure~\ref{fig:preprocessing_time_mass_v_mass} (bottom) shows the distribution of the duration that satellites spent in a preprocessing host halo.
The typical duration was $\sim 1.2 \gyr$, with 68\% spread of $0.5 - 3.5 \gyr$.
At $\mstar > 10 ^ 7 \msun$, no satellites were preprocessed longer than $1.8 \gyr$, while at $\mstar < 10 ^ 7 \msun$, the scatter increases significantly, with some satellites having experienced up to $7 \gyr$ of preprocessing.
The small durations at $\mstar > 10 ^ 7 \msun$ likely arise because those satellites have $\mpeak$ that approaches that of their preprocessing host halo, corresponding to shorter dynamical friction lifetimes, so they could not have been preprocessed too long without merging/disrupting.

Overall, most preprocessing occurred within groups of $\mvir = 10 ^ {10 - 12} \msun$, masses that feasibly could influence satellite galaxies, though environmental effects at these masses remain poorly understood.
Furthermore, the typical preprocessing duration was $0.5 - 3.5 \gyr$, comparable to typical timescales over which such satellite dwarf galaxies are quenched environmentally \citep{Fillingham2015, Wetzel2015b}.
Thus, we conclude that such group preprocessing before joining the MW/M31 halo is likely an important component in the evolution of satellite dwarf galaxies.

\renewcommand{\figurewidth}{0.99}
\begin{figure}
\centering
\includegraphics[width = \figurewidth \columnwidth]{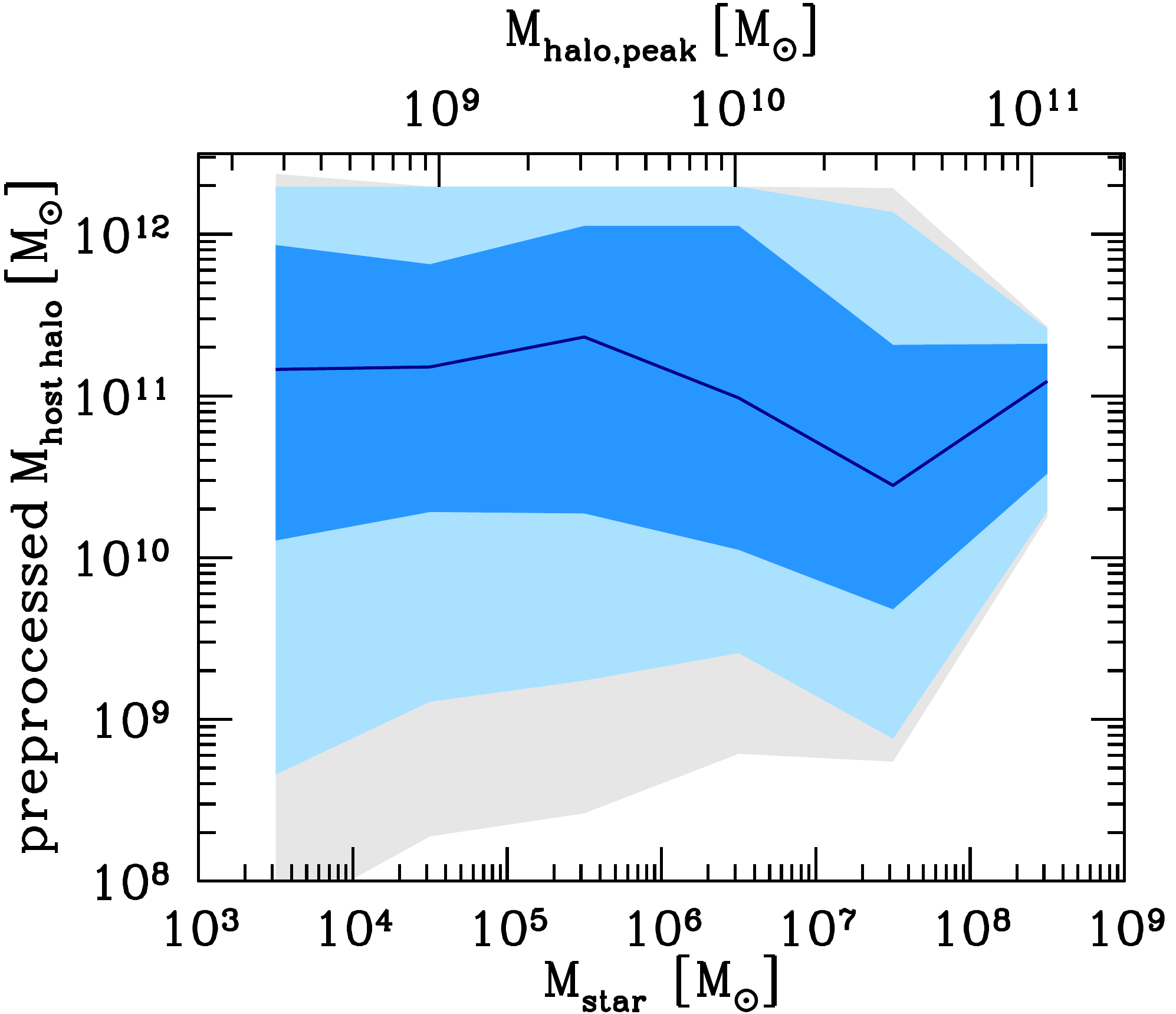}
\includegraphics[width = \figurewidth \columnwidth]{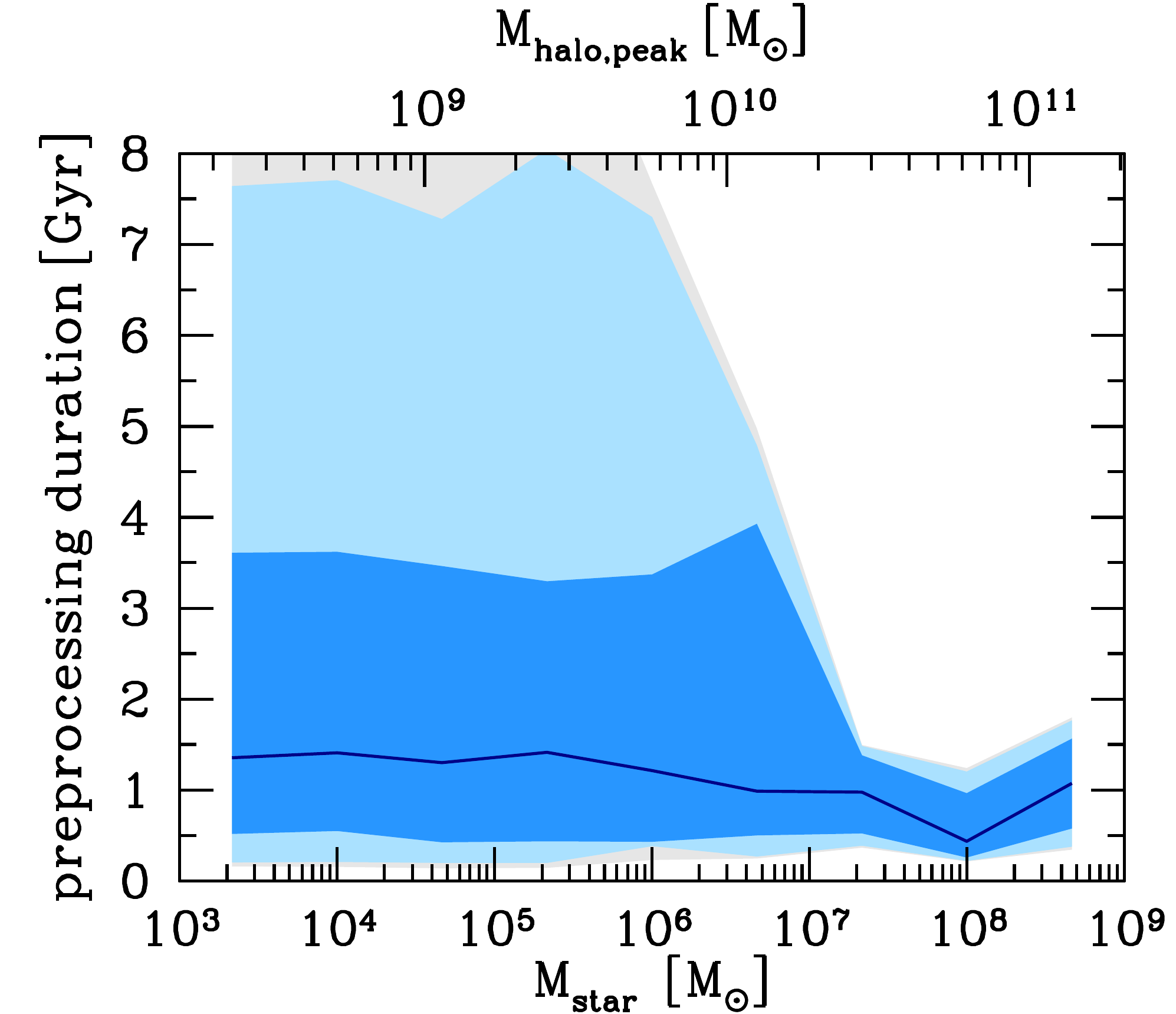}
\caption{
For satellite dwarf galaxies that were preprocessed in a group before falling into the MW/M31 halo, the maximum host-halo mass experienced during preprocessing (top) and the total duration of preprocessing (bottom), as a function of satellite stellar mass, $\mstar$, or subhalo peak mass, $\mpeak$.
Solid curves shows median, shaded regions show 68, 95, 99.7\% of the distribution.
Most preprocessing occurred in groups with $\mvir = 10 ^ {10 - 12} \msun$, masses at which environmental effects are feasible, though poorly understood.
The typical group preprocessing duration was $0.5 - 3.5 \gyr$, though some satellites at $\mstar < 10 ^ 7 \msun$ experienced significantly longer duration.
}
\label{fig:preprocessing_time_mass_v_mass}
\end{figure}

\section{Group Infall Drives Satellite-Satellite Mergers}
\label{sec:mergers}

\begin{figure}
\centering
\includegraphics[width = 0.99 \columnwidth]{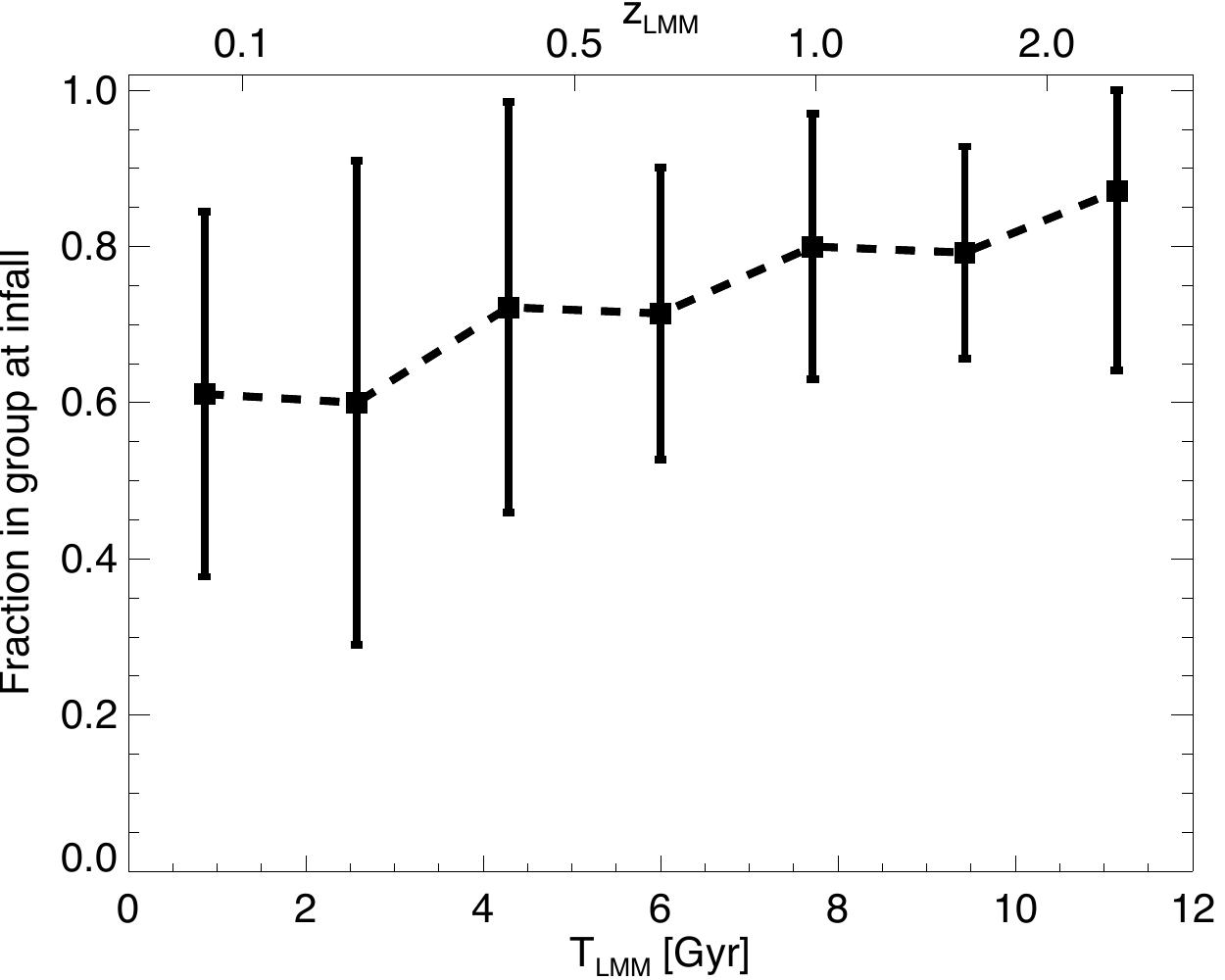}
\caption{
For all satellite galaxies at $z = 0$ with $\mstar = 10 ^ {3 - 9} \msun$ that experienced a major merger with another satellite after falling into the MW/M31 halo, the fraction of such mergers that fell in as part of the same host halo (group), as a function of the time since (or redshift of) the last major merger, $T_{\rm LMM}$ (or $z_{\rm LMM}$).
We do not find any significant dependence on satellite mass.
Infalling groups drive the majority ($60 - 90\%$) of all satellite-satellite mergers.
}
\label{fig:mergers}
\end{figure}

In \citet{Deason2014a}, we showed that most ($> 70\%$) satellites in the LG experienced a major merger ($\mstar$ ratio $> 0.1$) during their history.
While most mergers occurred prior to falling into the MW/M31 halo, a significant fraction were a satellite-satellite merger after infall.
Almost all of the latter occurred soon after falling in the MW/M31 halo \citep[see Figure 3 in][]{Deason2014a}, which suggests that such mergers occurred between satellites with correlated infall histories \citep[for example,][]{LiHelmi2008, Angulo2009, Wetzel2009a, Wetzel2009b}, in particular, that were part of the same group at infall.
We now demonstrate that group infall drives most such satellite-satellite mergers.

Figure~\ref{fig:mergers} shows, for all satellites that experienced a major merger after falling into the MW/M31 halo, what fraction occurred between galaxies that fell into the MW/M31 halo in the same group, as a function of the time since the last major merger, $T_{\rm LMM}$.
To address the limited statistics of major mergers, unlike in the rest of this paper, we combine all 48 paired and isolated MW/M31 halos in ELVIS, and we bin all satellites across $\mstar = 10 ^ {3 - 9} \msun$, given that we do not find any significant dependence on satellite mass.

The majority ($60 - 90\%$) of all satellite-satellite major mergers occurred between two galaxies that were in the same group when they fell into the MW/M31 halo.
This group infall drives a lower fraction of mergers at later cosmic time.
This likely relates to the larger delay time between MW/M31 infall and merging at later cosmic time \citep{Deason2014a}, which suggests that satellites had more opportunity to experience a ``chance'' merger with another satellite in the MW/M31 halo at later cosmic time.

Overall, in addition to driving environmental preprocessing, group infall is also an important catalyst for major mergers between satellite galaxies after they fall into the MW/M31 halo.

\section{Summary and discussion}
\label{sec:summary_discussion}

\subsection{Summary}
\label{sec:summary}

Using the ELVIS suite of cosmological zoom-in dissipationless simulations, we examined the virial-infall histories and group preprocessing of satellites across the observed range of masses for dwarf galaxies: $\mstar = 10 ^ {3 - 9} \msun$.
While we examined all 48 MW/M31 halos in ELVIS, we focused on the 10 halo pairs (20 halos total) that most resemble the LG, thus providing good statistics in a realistic cosmic setting.
We summarize our primary results as follows.

\begin{enumerate}
\renewcommand{\labelenumi}{(\alph{enumi})}
\item \textit{Virial-infall histories}: satellites at $z = 0$ fell into the MW/M31 halo typically $5 - 8 \gyr$ ago at $z = 0.5 - 1$, though they first fell into any host halo typically $7 - 10 \gyr$ ago at $z = 0.7 - 1.5$.
The difference between these infall times arises because of group preprocessing.
Satellites at lower mass or closer to the center of the MW/M31 experienced earlier infall times.
The latter means that the distribution of distances of satellites at $z = 0$ provides a statistical proxy for an environmental evolutionary sequence after infall.
Overall, current satellites have evolved as satellites within a host halo for over half of their entire history.

\item \textit{Group preprocesssing}: a large fraction of satellites were a satellite in a group (another host halo), typically of $\mvir \sim 10 ^ {10 - 12} \msun$, for a duration of $0.5 - 3.5 \gyr$, before falling into the MW/M31 halo.
This group preprocessing is especially common among faint and ultra-faint satellites: at $\mstar \lesssim 10 ^ 6 \msun$, $\approx 30\%$ of all satellites at $z = 0$ fell into the MW/M31 halo as a satellite in a group, and \textit{half} of all satellites at $z = 0$ were in a group any time before falling into the MW/M31 halo.
Thus, $\sim 25 \%$ of all satellites at $z = 0$ were preprocessed in groups with $\mvir \gtrsim 10 ^ {11} \msun$, comparable to the LMC.
Satellites closer to the center of the MW/M31 are more likely to have experienced group preprocessing.

\item \textit{Satellite-satellite mergers}: group infall drives most (60 - 90\%) of the satellite-satellite major mergers that occurred after falling into the MW/M31 halos, as we explored in \citet{Deason2014a}.

\item \textit{Cosmic reionization:} \textit{none} of the surviving satellite dwarf galaxies were within their MW/M31 halo during reionization ($z > 6$), and only $< 4\%$ were within the virial radius of any host halo during reionization.
Furthermore, the typical distance to the nearest, more massive halo at $z > 6$ was $3 \mpc$ comoving ($\sim 400 \kpc$ physical), or $500 \, \rvir$.
Thus, the effects of cosmic reionization versus host-halo environment on the formation histories of surviving dwarf galaxies in the LG occurred at distinct epochs separated typically by $2 - 4 \gyr$, and are separable in time theoretically and, in principle, observationally.
\end{enumerate}

\subsection{Discussion}
\label{sec:discussion}

\subsubsection{Impact of group preprocessing on the evolution of dwarf galaxies}

The significant fraction of satellite dwarf galaxies that experienced group preprocessing may help to explain the near-unity fraction of observed satellites in the LG that show strong environmental influence: spheroidal morphology, little-to-no cold gas, and quiescent star formation.
However, this depends on the extent to which the environmental processes, described in the Introduction, operate within low-mass groups of $\mvir \ll 10 ^ {12} \msun$.
If there is a lower limit in virial mass below which host halos do not significantly affect their satellites, then group preprocessing, even if common, many not be a particularly important regulator of the evolution of dwarf galaxies.
Because such low-mass groups necessarily are composed of faint galaxies, few observational works have probed the detailed properties of the satellites of such systems, though there are now ongoing efforts \citep[for example,][]{Stierwalt2014}.
Additionally, few theoretical works have examined the details of environmental effects in such lower-mass host halos, for example, in hydrodynamic simulations.
Based on our results, these would be fruitful areas for future investigation.

\subsubsection{Implications for observed associations in the Local Group}

The significant group-infall fractions (30 - 60 \%) that we found may help to explain the many observed associations between satellite galaxies (and stellar streams) within the halos of the MW and M31.
\citet{LiHelmi2008} showed that infalling groups can remain coherent and share similar orbital planes for up to $\sim 8 \gyr$ \citep[see also,][]{Klimentowski2010, Sales2011, SlaterBell2013}, which we find is also the typical time that satellites have been within the MW/M31 halo.
See also \citet{Deason2015}, for detailed phase-space distributions at $z = 0$ of infalling LMC-mass groups.

Observational evidence for associations between dwarf galaxies in the LG dates back to \citet{LyndenBell1982}, who demonstrated that the satellites in the MW halo appear situated along two great ``streams'': the Magellanic stream and the Fornax-Leo-Sculptor stream.
The discovery of faint and ultra-faint ($L \lesssim 10 ^ 5 L_\odot$) galaxies around the MW \citep{Willman2005, Belokurov2006, Belokurov2007a} led to further evidence for such galactic associations.
For example, the two ultra-faint galaxies Leo IV and Leo V are separated by $\sim 3$ degrees on the sky with small offsets in both distance and velocity \citep[see][for more discussion of such pairs]{Fattahi2013}.
Moreover, many works continue to explore (and debate) the presence of a planar, disk-like distribution of satellites around both the MW and M31 \citep[for example,][]{Libeskind2005, Deason2011, Lovell2011, Fattahi2013, Ibata2013, PawlowskiKroupa2014}, which could be the result of one or more infalling groups.
Note, however, that \citet{PawlowskiMcGaugh2014} examined the satellite configurations in ELVIS and found no polar structures analagous to that around the MW.
Our results also fully support the likelihood that the SMC was in a group with the LMC prior to MW infall \citep{Kallivayalil2013}.

In addition to associations between galaxies, many authors have noted observed association between galaxies, stellar streams, and/or structures in the stellar halo.
For example, \cite{Newberg2010} showed that the Orphan stellar stream \citep{Grillmair2006, Belokurov2007b} has a similar distance, velocity, and position as Segue 1.
Likewise, the proximity of Bo{\"o}tes II in both position and velocity to the Sagittarius stream led \citet{Koch2009} to suggest that Bo{\"o}tes II may have been stripped from the more massive Sagittarius dwarf galaxy.
In addition, \citet{Deason2014b} \citep[see also][]{Belokurov2009} showed that Segue 2, which is perhaps the least-massive known galaxy \citep{Kirby2013a}, is likely associated with the large, metal-rich Triangulum-Andromeda overdensity, and \citet{Kirby2013b} showed that Segue 2 lies off of the tight mass-metallicity relation for most dwarf galaxies, which may indicate of group infall.
We reiterate that in about half of our cases of group infall, the lower-mass satellite from the group survives to $z = 0$ in the MW/M31 halo, while the more massive (primary) galaxy merges/disrupts, which could lead to such observable associations between (surviving) galaxies and (disrupted) stellar streams.
Overall, the enhanced evidence for associations between the lowest-mass satellites agrees well with our predictions (Figure~\ref{fig:infall.fraction_v_mass}).

\subsubsection{Effects of numerical resolution and baryonic physics}

For any simulation, the survivable lifetime of a satellite depends on how well the simulation resolves it, which can lead to prematurely merging/disrupting (as compared with a real galaxy),  especially for lower-mass satellites and/or those that fell in at high redshift.
If this were a strong numerical effect in ELVIS, then the virial-infall times for surviving satellites at $z = 0$ would be biased to lower redshifts.
However, using the three isolated MW/M31 halos in ELVIS that were re-run at $2 \times$ higher spatial and $8 \times$ higher mass resolution, we checked that the virial-infall times do change change significantly in these higher-resolution runs.
This agrees with the resolution tests in \citet{GarrisonKimmel2014}, which demonstrated the completeness of the satellite population at our mass range, $\mpeak > 10 ^ 8 \msun$.

A more significant concern is that ELVIS simulates only the gravitational dynamics of dark matter, and baryonic effects may change the survivability and stellar content of subhalos, in at least two ways.
First, we assumed that all subhalos with $\mpeak > 10 ^ 8 \msun$ host luminous galaxies, according to abundance matching against $\mpeak$, regardless of when the subhalos formed.
We emphasize that this approach is largely consistent with the observed mass function of satellites in the LG \citep{GarrisonKimmel2014}, especially if one accounts for observational incompleteness \citep{Tollerud2008, Hargis2014}.
Furthermore, even if not all subhalos host luminous galaxies in reality, this would not bias our results if luminous galaxies populate subhalos largely stochastically.
However, some results from cosmological baryonic simulations \citep[for example,][]{Sawala2014} suggest that the subhalos that do host luminous galaxies are the ones that formed preferentially earlier, when they had deeper potential wells.
If true, then this baryonic effect would shift the virial-infall times of surviving, luminous satellites to have occurred at higher redshifts.

Second, the addition of the baryonic disk of the host MW/M31 galaxy can lead to more rapid disruption of satellites through tidal shocking or resonant stripping \citep{Mayer2001, DOnghia2010, Zolotov2012}.
If a strong effect, then our dark-matter simulations would overestimate the survival lifetimes of satellites after infall, and the virial-infall times of surviving satellites would need to shift to lower redshifts.

Thus, the combination of all potential baryonic effects could shift our virial-infall times in either direction, and future work should elucidate these trends with statistical samples of baryonic simulations.
However, we do not expect that these baryonic effects would alter our results regarding reionization.
\\

After our submission, \citet{Koposov2015}, \citet{Bechtol2015}, and \citet{Martin2015} announced the discovery of multiple faint dwarf galaxies near the LMC.
Our results strongly support the likelihood that many of these were satellites of the LMC prior to MW infall.
\\

We thank the Aspen Center for Physics, supported in part by the National Science Foundation, for the hospitality and stimulating environment during the preparation of this paper.
We thank Erik Tollerud, Dan Weisz, and Laura Sales for useful discussions.
We also thank the anonymous reviewer for useful comments.
A.R.W. gratefully acknowledges support from the Moore Center for Theoretical Cosmology and Physics at Caltech.
A.J.D. is supported by NASA through Hubble Fellowship grant HST-HF-51302.01, awarded by the Space Telescope Science Institute, which is operated by the Association of Universities for Research in Astronomy, Inc., for NASA, under contract NAS5-26555.

\appendix

\section{Satellites in Paired Versus Isolated Host Halos}

\renewcommand{\figurewidth}{0.48}
\begin{figure*}
\centering
\includegraphics[width = \figurewidth \textwidth]{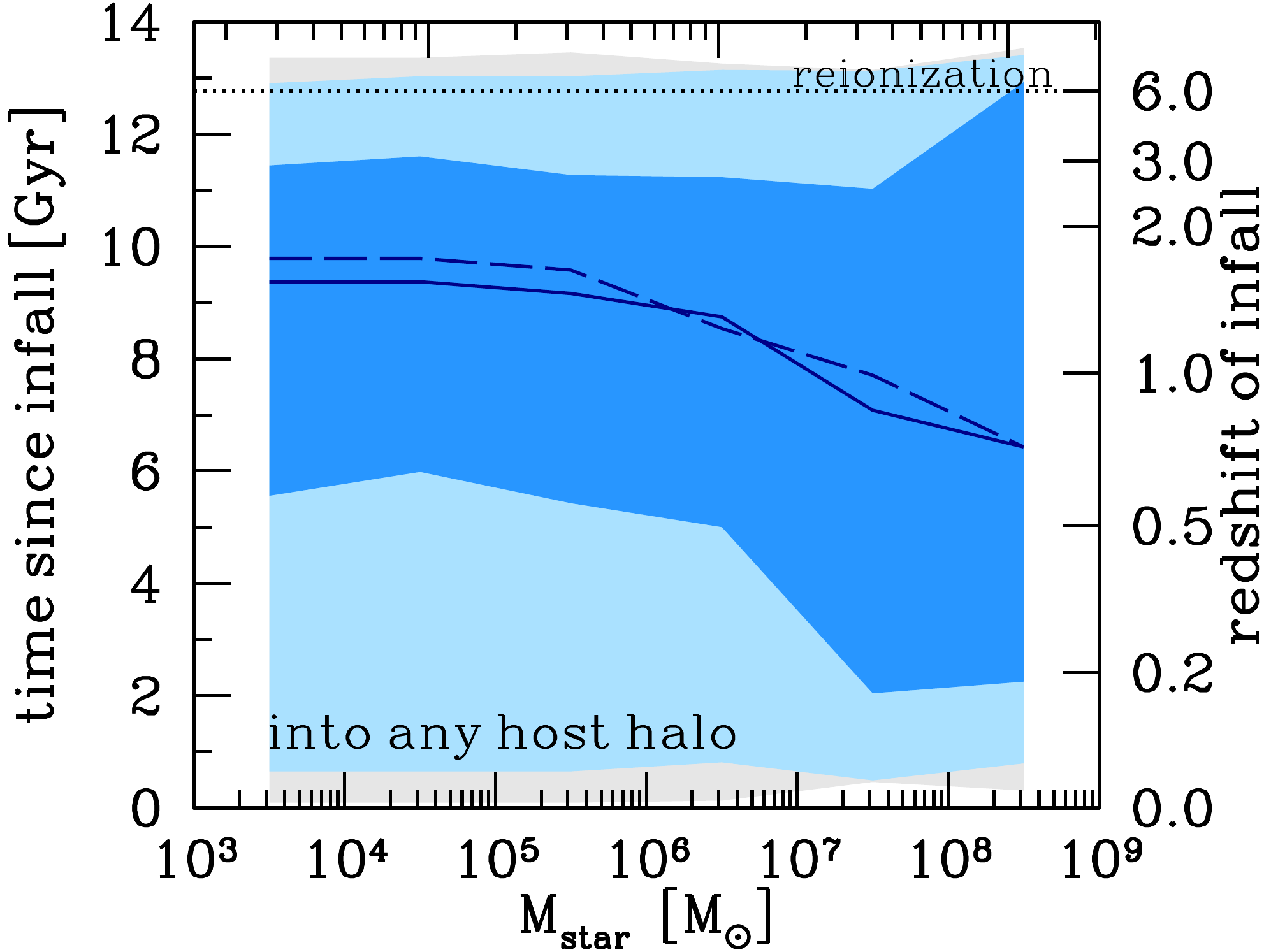}
\hspace{3 mm}
\includegraphics[width = \figurewidth \textwidth]{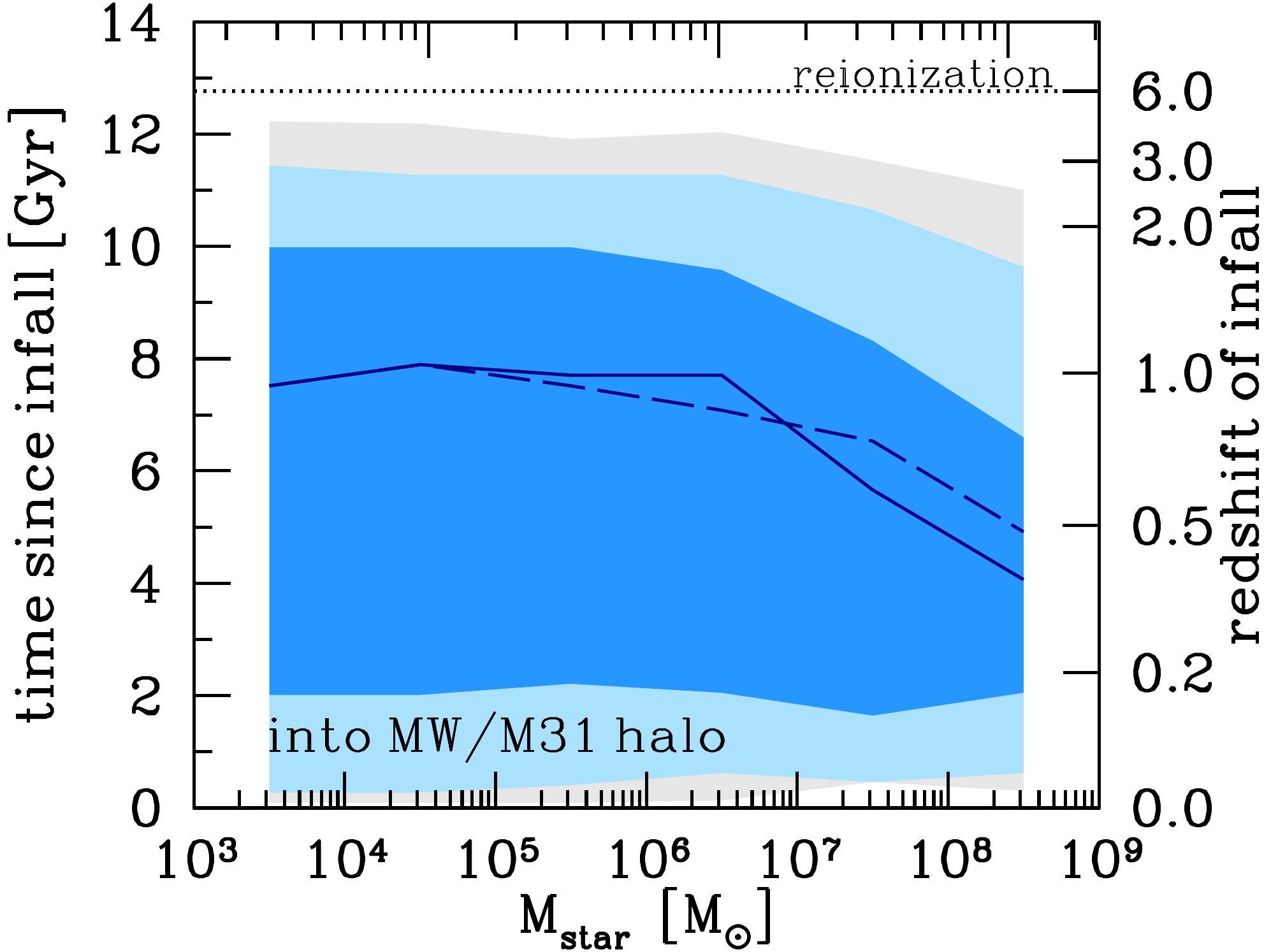}
\caption{
Same as Figure~\ref{fig:infall.time_v_mass}, but for satellite dwarf galaxies in the 24 isolated MW/M31 halos, which are matched in mass to the LG-like paired MW/M31 halos.
For comparison, dashed curves show median values from the paired halos from Figure~\ref{fig:infall.time_v_mass}.
We find no significant differences in the virial-infall times of satellites in the isolated versus paired halos.
}
\label{fig:infall.time_v_mass_isolated}
\end{figure*}

\begin{figure}
\centering
\includegraphics[width = 0.99 \columnwidth]{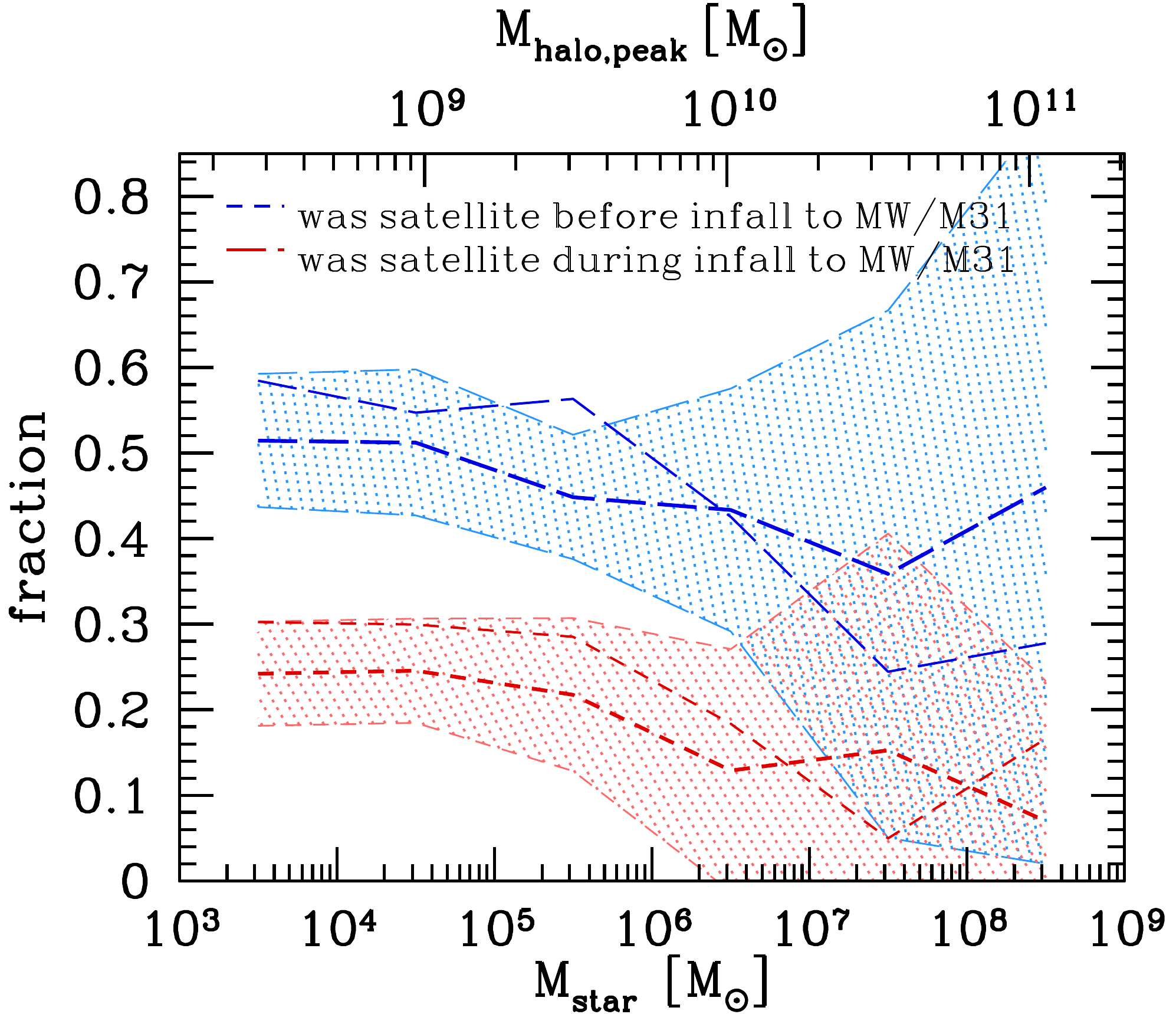}
\caption{
Same as Figure~\ref{fig:infall.fraction_v_mass}, but for satellite dwarf galaxies in the isolated MW/M31 halos, which are matched in mass to the LG-like paired MW/M31 halos.
For comparison, lighter dashed curves show averages from the paired halos from Figure~\ref{fig:infall.fraction_v_mass}.
For low-mass satellites, those in the isolated halos are less likely to have been preprocessed in a group, at a level that is comparable to the halo-to-halo standard deviation.
For high-mass satellites, any differences are well within the standard deviation.
}
\label{fig:infall.fraction_v_mass_isolated}
\end{figure}

While thus far we presented results using just the paired MW/M31 halos in ELVIS, here we compare our main results---virial-infall times and group preprocessed fractions---for satellites in the isolated versus paired MW/M31 halos.
This comparison is useful for a number of reasons.
First, theoretically, we want to understand the degree to which the larger-scale environment around a MW/M31 halo influences the infall histories of its satellite population.
Second, this comparison informs whether theoretical models need to consider separately the satellite populations of paired versus isolated MW/M31 halos, for example, in order to understand satellites in the LG versus in isolated MW/M31-like galaxies in the local volume.

Figure~\ref{fig:infall.time_v_mass_isolated} shows $\tsinceinfallfirst$ (left) and $\tsinceinfallhost$ (right) as a function of satellite $\mstar$ or $\mpeak$, similar to Figure~\ref{fig:infall.time_v_mass}, but for the isolated MW/M31 halos.
For comparison, dashed curves show the median values for the paired MW/M31 halos from Figure~\ref{fig:infall.time_v_mass}.
While satellites in isolated MW/M31 halos first fell into a host halo slightly later, any such difference is small compared to the scatter.
Thus, we conclude that the virial-infall times of satellites do not depend significantly on whether their current host halo is isolated or paired like in the LG.

Similar to Figure~\ref{fig:infall.fraction_v_mass}, Figure~\ref{fig:infall.fraction_v_mass_isolated} shows the fraction of all satellites at $z = 0$ that were a satellite in another host halo any time before falling into the MW/M31 halo (blue long dashed), or at the time of falling into the MW/M31 halo (red short dashed), as a function of satellite $\mstar$ or $\mpeak$.
For comparison, the lighter dashed curves show the averages for the paired MW/M31 halos from Figure~\ref{fig:infall.fraction_v_mass} (left).
Here, differences between isolated versus paired MW/M31 halos are stronger, such that low-mass satellites are more likely to have been preprocessed if they are in paired MW/M31 halos, at a level comparable to the halo-to-halo standard deviation (shaded region).
This trend reverses slightly at higher mass, but here the difference is much smaller than the scatter, so we do not consider it significant.

Most likely, the higher group-preprocessed fractions for satellites in the paired MW/M31 halos arises because, as \citet{GarrisonKimmel2014} noted, the paired halos have many more neighboring halos within a few Mpc of them than the isolated halos, because the paired halos (almost by definition) reside in a preferentially higher-mass cosmic region.
With more neighboring halos around, the satellites in the paired MW/M31 halos are more likely first to have fallen into a neighboring host halo.
However, this difference in preprocessed fraction does not lead to a significant difference in infall times (Figure~\ref{fig:infall.fraction_v_mass_isolated}), so while group preprocessing is more prevalent for satellites in the paired MW/M31 halos, the duration of this preprocessing is not longer.

Finally, the most significant difference that we find between the satellites in isolated versus paired MW/M31 halos was in Figure~\ref{fig:nearest_distance_reionization}: during the epoch of reionization ($z > 6$), the progenitors of the satellites in the isolated MW/M31 halos were much ($\sim 2 \times$) closer to their nearest neighboring, more massive halo than those in the paired MW/M31 halos.
This result may seem counterintuitive, given that the paired halos contain many more neighboring halos at $z = 0$.
However, we find that these structures were diluted over a much larger volume at $z > 6$ for the paired halos.
Specifically, we randomly sub-sample all particles within $\rvir$ of each MW/M31 halo at $z = 0$ and trace their locations back to $z > 6$, finding that the Lagrangian volume that contains all such particles was many ($2 - 6$) times larger for the paired MW/M31 halos, such that the satellite progenitors from the paired MW/M31 halos had fewer neighboring halos at a given distance at $z > 6$.


\end{document}